\documentclass{article}

\usepackage{amsmath,amstext,amssymb,epsf}
\usepackage{epsfig}
\numberwithin{equation}{section} % in amsmath

 \newtheorem{lemma}{Lemma}[section]
 \newtheorem{theorem}[lemma]{Theorem}

 \newtheorem{definition}[lemma]{Definition}
 \newtheorem{rem}[lemma]{Remark}
\newenvironment{remark}{\begin{rem}}{\hspace*{\fill}$\diamondsuit$\end{rem}}
\newenvironment{proof}{\par \sc Proof.\rm}{\hspace*{\fill}$\Box$\vspace{1ex}}
 \newtheorem{ex}[lemma]{Example}

\newcommand{\NCD}{\textsc {NCD} }

\newcommand{\commentout}[1]{}

\begin{document}

\title{A New Quartet Tree Heuristic for Hierarchical Clustering}
\author{Rudi Cilibrasi and Paul M.B. Vit\'{a}nyi
\thanks{
Rudi Cilibrasi is with the Centre for Mathematics
and Computer Science (CWI). Address:
CWI, Kruislaan 413,
1098 SJ Amsterdam, The Netherlands.
Email: {\tt Rudi.Cilibrasi@cwi.nl}.
 Part of his work was supported by
 the Netherlands BSIK/BRICKS project, and by NWO  project 612.55.002.
Paul Vit\'{a}nyi is with the Centre for Mathematics and Computer Science (CWI),
and the University of Amsterdam.
Address:
CWI, Kruislaan 413,
1098 SJ Amsterdam, The Netherlands.
Email: {\tt Paul.Vitanyi@cwi.nl}.
He was supported in part by the
EU project RESQ, IST--2001--37559, the NoE QUIPROCONE
IST--1999--29064,
the ESF QiT Programmme, and the EU NoE PASCAL,
the Netherlands BSIK/BRICKS project.
}}

%\markboth{IEEE Transactions on Information Theory, VOL. XX, NO Y, MONTH 2005}{Rudi Cilibrasi and Paul Vit\'{a}nyi: Clustering by Compression}

\maketitle

\begin{abstract}
We consider the problem of constructing an 
an optimal-weight tree
from the $3{n \choose 4}$ weighted quartet topologies on $n$ objects, 
where optimality means that the summed weight of the embedded quartet topologies
is optimal (so it can be the case that the optimal tree embeds all quartets
as non-optimal topologies).
We present a heuristic for reconstructing the optimal-weight tree,
and a canonical manner
to derive the quartet-topology weights from a given distance matrix. The method
repeatedly transforms a bifurcating tree, 
with all objects involved as leaves, achieving 
a monotonic approximation to the exact single globally optimal tree.
This contrasts to other heuristic search methods 
from biological phylogeny, like DNAML
or quartet puzzling,
which, repeatedly, incrementally
construct a solution from a random order of objects,
and subsequently add agreement values.
We do not assume that there exists a true bifurcating supertree that
embeds each quartet in the optimal topology, or represents the distance
matrix faithfully---not even under the assumption that the weights or distances
are corrupted by a measuring process. 
Our aim is to hierarchically cluster the input data as faithfully as possible,
both phylogenetic data and data of completely different types.
In our experiments with natural data, like genomic data, texts or music, the global
optimum appears to be reached.
Our method is capable of handling over 100
objects, possibly up to 1000 objects, while no existing quartet heuristic
can computionally approximate the exact optimal solution of a quartet tree 
of more than about 20--30 objects
without running for years.
The method is implemented and available as public software.
\end{abstract}

\section{Introduction}
\label{sect.intro}

We present a method of hierarchical clustering based on
a novel fast randomized hill-climbing heuristic of a new global 
optimization criterion. Given a the weights of all quartet topologies,
or a matrix of the pairwise
distances between the objects, we obtain an output tree with the objects
as leaves, and we score how well the tree
represents the information in the distance matrix on a scale of 0 to 1.
As proof of principle, we experiment on three
data sets, where we know what
the final answer should be: (i) reconstruct a tree from a distance
matrix obtained from a randomly generated tree; (ii)
reconstruct a tree from files containing artificial similarities;
and (iii) reconstruct a tree from natural files of
heterogenous data of vastly different types.
We give examples in whole-genome phylogeny
using the whole mitochondrial DNA of the species concerned, in SARS virus
localization among other virri, and in analyzing the spreading of the bird-flu 
H5N1 virus mutations.
We compare the hierarchical clustering of our method with
a more standard method of two-dimensional clustering (to show
that our dendrogram method of depicting the clusters is more informative).
The new method was developed as an auxiliary tool
for \cite{CVW03,CV04}, since the available
quartet tree methods were too slow 
when they were exact, and too inaccurate or uncertain 
when they were statistical incremental.
Our new quartet tree heuristic
runs orders of magnitudes faster than any other exact quartet tree
method,
and gives consistently good results in practice.

{\bf Relation with Previous Work:}
The  Minimum Quartet Tree Cost (MQTC) problem below for which we
give a new computational heuristic is related to
the Quartet Puzzling problem,
\cite{SvH96}. There, the quartet topologies are provided
with a probability value, and for each quartet the topology with
the highest probability is selected (randomly, if there are more than one)
as the maximum-likelihood optimal
topology. 
The goal is to find a bifurcating tree
that embeds these optimal quartet
topologies. In the biological setting it is assumed that the observed genomic data 
are the result of an evolution in time, and hence can be represented as the leaves
of an evolutionary tree. Once we obtain a proper probabilistic evolutionary
model to quantify the evolutionary relations between the data we can search  
for the true tree. In a quartet method one determines the most likely quartet
topology under the given assumptions, and then searches for a tree that represents
as many of such topologies as is possible. If the theory and data were perfect
then there was a tree that represented precisely all most likely quartet topologies.
Unfortunately, in real life the theory is not perfect, the data are corrupted,
and the observation polutes and makes errors. Thus, one has to settle for
embedding as many most likely quartet topologies as possible, do error correction
on the quartet topologies, and so on. For $n$ objects, there are 
$(2n-5)!! = (2n-5) \times  (2n-3) \times \cdots \times 3$ unrooted bifurcating trees.
For $n$ large, exhaustive search for the optimal tree is impossible,
and turns out to be
NP-hard, and hence infeasible in general. There are two main avenues that have
been taken:

(i) Incrementally grow the tree in random order by stepwise addition
of objects in the current optimal way, repeat this for different
object orders, and add agreement values on the branches, like DNAML
\cite{Fe81}, or quartet puzzling \cite{SvH96}.
 
(ii) Approximate the global optimum monotonically or compute it,
using  geometric algorithm or
dynamic programming \cite{BCGOP98}, and linear programming \cite{WDGG05}.

These methods, other methods, as well as methods related to the MQT problem,
cannot handle more than 15--30 objects \cite{WDGG05,LTM05,PBE04,BJKLW99}
directly,
even while
using farms of desktops.
To handle more objects one needs to construct a supertree from the constituent
quartet trees for subsets of the original data sets, \cite{RMWW04}, as
in \cite{LTM05,PBE04}.

In 2003 in \cite{CVW03,CV04} we considered a new approach, like \cite{WDGG05}, and
possibly predating it. Our goal was to use a quartet method to obtain high-quality
hierarchical clustering of data from arbitrary (possibly heterogenous)
domains, not necessarily phylogeny data. We thus do not assume that there exists
a true evolutionary tree, and our aim is not to just embed as many 
optimal quartet topologies as is possible. Instead, for $n$ objects
we consider all $3{n \choose 4}$ possible quartet topologies, each with
a given weight, and our goal is to find the tree such that the summed weights
of the embedded quartet topologies is optimal. We develop an heuristic that
monotonically approximates this optimum, a figure of merit that quantifies
the quality of the best current candidate tree. We show that the problem is NP-hard,
but we give evidence that the natural data sets we consider have qualities of
smoothness so that the monotonic heuristic obtains the global optimum
in a feasible number of steps.

{\bf Materials and Methods:}
Some of the experiments reported are taken from \cite{CVW03,CV04}
where many more can be found.
The data samples we used were obtained from standard data bases
accessible on the world-wide web, generated by ourselves,
or obtained from research groups in the field of investigation.
We supply the details with each experiment.
The clustering heuristic generates a tree
with an optimality quantification, 
called standardized benefit score or $S(T)$ value
in the sequel. Contrary to other phylogeny methods,
we do not have agreement or confidence values on the branches:
we generate the best tree possible, globally balancing all requirements.
Generating trees from the same distance
matrix many times resulted in the same tree in case of high $S(T)$
value, or a similar tree in case of moderately high $S(T)$ value,
for all distance matrices we used, even though the heuristic is randomized.
That is, there is only one way to be right, but increasingly
many ways to be increasingly wrong which can all be realized by different
runs of the randomized algorithm.
The quality of the results depends on
how well the hierarchical tree represents the information
in the matrix. That quality is measured by the $S(T)$ value,
and is given with each experiment.
In certain natural data sets, such as H5N1 genomic sequences, consistently high
$S(T)$ values are returned even for large sets of objects of 100 or more nodes.
In other discordant natural data sets however, the $S(T)$ value deteriorates
more and more with increasing number of elements being put in the same tree.
The reason is that with increasing size of a discordant natural data set
the projection of the information in the distance matrix into a
ternary tree gets necessarily increasingly distorted because the underlying
structure in the data is incommensurate with any tree shape whatsoever.
In this way, larger structures may induce additional ``stress'' in the mapping
that is visible as lower and lower $S(T)$ scores.

{\bf Figures:}
We use two styles to display the hierarchical clusters.
In the case of genomics of Eutherian orders,
it is convenient to follow the
dendrograms that are customary in that area (suggesting temporal
evolution) for easy comparison with the literature.
In the other experiments
(even the genomic SARS experiment) it is more informative to
display an unrooted ternary tree (or binary tree if we think about
incoming and outgoing edges) with explicit internal nodes.
This facilitates identification of clusters in terms of
subtrees rooted at internal nodes or contiguous sets of subtrees rooted
at branches of internal nodes.

\section{Hierarchical Clustering}

Given a set of objects as points in a space provided with
a (not necessarily metric) distance measure,
the associated {\em distance matrix}
has as entries the pairwise distances
between the objects. Regardless of the original space and distance measure,
it is always possible to configure $n$ objects in $n$-dimensional
Euclidean space in such a way that the associated distances are
identical to the original ones, resulting in an identical distance
matrix.
This distance matrix contains the pairwise distance relations
according to the chosen measure in raw form. But in this format
that information is not easily usable, since for $n > 3$ our
cognitive capabilities rapidly fail.
Just as the distance matrix is a reduced form of information
representing the original data set, we now need to reduce the
information even further in order to achieve a cognitively acceptable
format like data clusters.
To extract a hierarchy of clusters
from the distance matrix,
we determine a dendrogram (ternary tree) that agrees
with the distance matrix according to a cost measure.
This allows us to extract more information from the data
than just flat clustering (determining disjoint
clusters in dimensional representation).

Clusters are groups of objects that are similar
according to our metric. There are various ways
to cluster. Our aim
is to analyze data sets for which the number of clusters is
not known a priori, and the data are not labeled. As stated in \cite{DHS},
conceptually simple, hierarchical clustering is among
the best known unsupervised methods in this setting, and
the most natural way is to represent the relations
in the form of a dendrogram, which is customarily a directed binary tree
or undirected ternary tree.
With increasing number of data items, the projection of the distance matrix
information into the tree representation format may get distorted.  Not all
natural data sets exhibit this phenomenon; but for some, the tree gets
increasingly distorted as more objects are added.  A similar situation
sometimes arises in using alignment cost in genomic comparisons.
Experience shows that in both cases the hierarchical clustering methods
seem to work best for small sets of data, up to 25 items, and to deteriorate
for some (but not all) larger sets, say 40 items or more.  This deterioration
is directly observable in the $S(T)$ score and degrades solutions in two
common forms: tree instability when different or very different solutions are returned on successive runs or tree ``overlinearization''
when some data sets produce caterpillar-like structures only or predominantly.
In case a large set of objects, say 100 objects, clusters with high $S(T)$
value this is evidence that the data are of themselves tree-like, and
the quartet-topology weights, or underlying distances, truely represent to
similarity relationships between the data.
%A standard solution to hierarchically cluster larger sets of data is to first
%cluster nonhierarchically, by say multidimensional scaling of $k$-means,
%available in standard packages, for instance {\em Matlab},
%and then apply hierarchical clustering on the emerging clusters.

\section{The Quartet Method}
\begin{figure}[htb]
\begin{center}
\epsfig{file=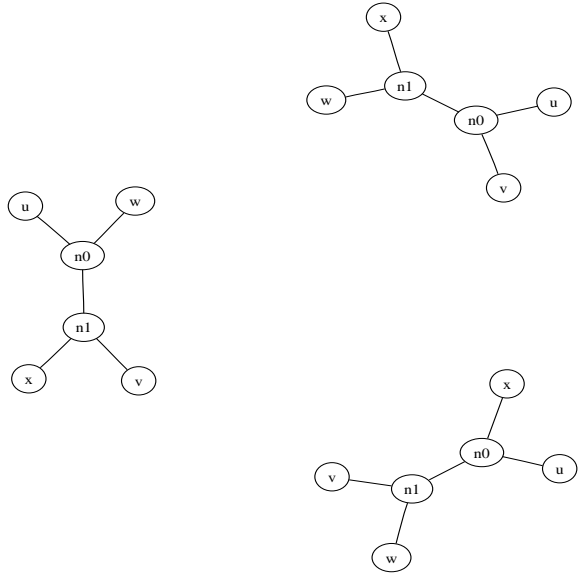,width=3in,height=3in}
\end{center}
\caption{The three possible quartet topologies for the set of leaf labels {\em u,v,w,x} }\label{figquart}
\end{figure}

Given a set $N$ of $n$ objects,
we consider every set of four elements from our set
of $n$ elements;
there are ${n \choose 4}$ such sets.
From each set $\{u,v,w,x\}$ we construct a tree of arity 3,
which implies that the tree consists of two subtrees of two
leaves each. Let us call such a tree a {\em quartet topology}.
The set of $3 {n \choose 4}$ quartet topologies induced by $N$
is denoted by $Q$.
We denote a partition $\{u,v\},\{w,x\}$ of $\{u,v,w,x\}$
by $uv | wx$.
There are
three possibilities to partition $\{u,v,w,x\}$ into
two subsets of two elements each: (i) $uv | wx$, (ii) $uw | vx$,
and (iii)  $ux | vw$.  In terms of the tree topologies:
 a vertical bar divides the two pairs of leaf nodes
into two disjoint subtrees (Figure~\ref{figquart}).

\begin{definition}
\rm
For the moment we consider the class ${\cal T}$ of undirected trees
of arity 3 with $n \geq 4$ leaves,
labeled with the elements of $N$.
\end{definition}
Such trees have $n$ leaves and $n-2$ internal nodes.
%This restriction and its possible relaxation
%will be discussed later.
For any given tree $T$ from this class, and any set
of four leaf labels $u,v,w,x \in N$, we say $T$ is $consistent$ with $uv | wx$
if and only if the path from $u$ to $v$ does not cross
the path from $w$ to $x$. It is easy to see that
precisely one of the three possible
quartet topologies for any set of 4 labels is consistent
for a given tree from the above class, and therefore a tree from ${\cal T}$
contains precisely ${n \choose 4}$ different quartet topolgies.
\begin{figure}[htb]
\begin{center}
\epsfig{file=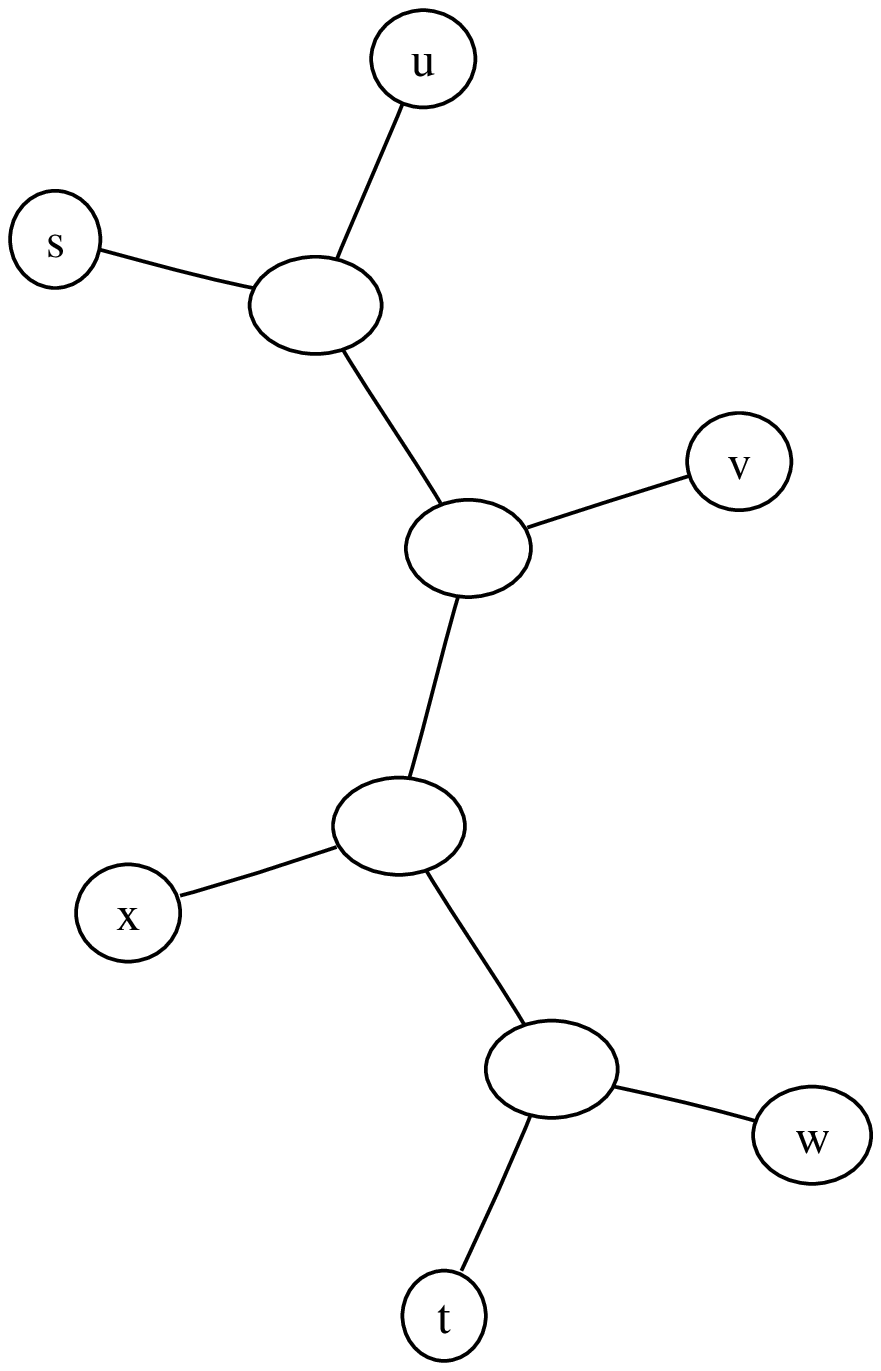,width=2in,height=3in}
\end{center}
\caption{An example tree consistent with quartet topology $uv | wx$ }\label{figexquart}
\end{figure}
We may think of a large tree having many smaller quartet topologies embedded
within its structure. Commonly the goal in the quartet method
is to find (or approximate as closely as possible) the tree
that embeds the maximal number of consistent (possibly weighted) quartet
topologies from a given set $P \subseteq Q$ of quartet topologies
\cite{Ji01} (Figure~\ref{figexquart}).
A  {\em weight function} $W: P \rightarrow {\cal R}$, with ${\cal R}$
the set of real numbers determines the weights. The unweighted case is
when $W(uv|wx)=1$ for all $uv|wx \in P$.
\begin{definition}
\rm
The (weighted) {\em Maximum Quartet Consistency (MQC)}
is defined as follows:

GIVEN: $N$, $P$, and $W$.

QUESTION: Find $T_0 = \max_T  \sum \{ W(uv|wx): uv|wx \in P$
and $uv|wx$ is consistent
with $T \}$.
\end{definition}

\section{Minimum Quartet Tree Cost}
The rationale for the MQC optimization problem is the assumption that
there is exists a tree $T_0$ as desired in the class ${\cal T}$
under consideration,
and our only problem is to find it. This assumption reflects the
genesis of the method in the phylogeny community. Under the
assumption that biological species developed by evolution in time, and $N$ is
a subset of the now existing species, there is a phylogeny $P$ (tree in
${\cal T}$) that represents that evolution. The set of quartet topologies
consistent with this tree, has one quartet topology per quartet which is
the true one. The quartet topologies in $P$ are the ones which we assume
to be among the true quartet topologies, and weights are used to express
our relative certainty about this assumption concerning the individual
quartet topologies in $P$.

However, the data may be corrupted so that this assumption is no longer true.
In the general case of hierarchical clustering we do not even have a priori
knowledge that certain quartet topologies are objectively
true and must be embedded.
Rather, we are in the position that we can somehow assign a relative importance
to the different quartet topologies. Our task is then to balance the
importance of embedding different quartet topologies against one another,
leading to a tree that represents the concerns as well as possible.
%The most widely used heuristic of this type is {\em Quartet Puzzling},
%specifically oriented to biological
%phylogeny, \cite{SvH96}, where the weights are taken from a posterior
%maximum likelihood distribution. 
%Below we review the relevant theory and
%give formal details where they are lacking or folklore in the literature.
%Consider a slight generalization of Quartet Puzzling, where 
We start
from a cost-assignment to the quartet topologies; the method by which we
assign costs to the $3{n \choose 4}$ quartet topologies is
for now immaterial to our problem.
Given a set $N$ of $n$ objects, let $Q$ be the set of
quartet topologies, and let $C:Q \rightarrow {\cal R}$ be a {\em cost function}
assigning a real valued cost $C_{uv|wx}$ to each quartet
$uv|wx \in Q$.
\begin{definition}
The {\em cost} $C_T$ of a tree $T$ with a set $N$ of leaves (external nodes
of degree 1) is defined by
$C_T =\sum_{\{u,v,w,x\} \subseteq N} \{C_{uv|wx}: T$ is consistent
with  $uv |wx\}$---the
sum of the costs of all its consistent quartet topologies.
\end{definition}

\begin{definition}
\rm
Given $N$ and $C$,
the {\em Minimum Quartet Tree Cost (MQTC)} is
$\min_T \{C_T:$ $T$ is a tree with
the set $N$ labeling its leaves$\}$.
\end{definition}

We normalize the problem of finding the MQTC as follows:
Consider the list of all possible quartet topologies
for all four-tuples of labels
under consideration.  For each group of
three possible quartet topologies for a given
set of four labels $u,v,w,x$, calculate a best (minimal) cost
$m(u,v,w,x) = \min \{ C_{uv|wx}, C_{uw|vx}, C_{ux|vw} \}$,
and a worst (maximal)
cost $M(u,v,w,x) = \max \{ C_{uv|wx}, C_{uw|vx}, C_{ux|vw} \}$.
Summing all best quartet toplogies yields the best (minimal) cost
$m = \sum_{\{u,v,w,x\} \subseteq N} m(u,v,w,x)$.
Conversely, summing all worst quartet toplogies yields the worst (maximal) cost
$M =  \sum_{\{u,v,w,x\} \subseteq N} M(u,v,w,x)$.
For some distance matrices,
these minimal and maximal values can not be attained by actual trees;
however, the score $C_T$ of every tree $T$ will lie between these two values.
In order to be able to compare the scores of quartet trees for different
numbers of objects in a uniform way,
we now rescale the score linearly such that the worst score maps to 0,
and the best score maps to 1:

\begin{definition}
\rm
The {\em normalized tree benefit score} $S(T)$ is defined
by $S(T) = (M-C_T)/(M-m)$.
\end{definition}

Our goal is to find a full tree with a maximum value
of $S(T)$, which is to say, the lowest total cost.
Now we can rephrase the MQTC problem in such a way that
solutions of instances of different sizes can be
uniformly compared in terms of relative quality:

\begin{definition}
\rm
Definition of the {\em MQTC problem}:

GIVEN: $N$ and $C$.

QUESTION: Find a tree $T_0$ with $S(T_0)=\max \{S(T):$ $T$ is a tree with
the set $N$ labeling its leaves$\}$.
\end{definition}

\subsection{Computational Hardness}
The hardness of Quartet Puzzling is informally mentioned in the literature
\cite{WDGG05,LTM05,PBE04}, but we provide explicit proofs.
To express the notion of computational difficulty one uses
the notion of ``nondeterministic polynomial time (NP)''.
If a problem concerning $n$ objects is NP-hard
this means that the best known algorithm
for this (and a wide class of significant problems) requires
computation time exponential in $n$. That is, it is infeasible
in practice.
The {\em MQC decision problem} is the following:
Given a set $N$ of $n$ objects, let $T$ be a tree of which the $n$
leaves are labeled by the objects, and let $Q$ be the set of quartet topologies
and $Q_T$ be the set of quartet topologies
embedded in $T$.
Given a set of quartet topologies $P \subseteq Q$,
and an integer $k$,
the problem is to decide whether there is
a binary tree $T$ such that $P \bigcap Q_T > k$.
In \cite{St92} it is shown that the MQC decision problem
is NP-hard. We have formulated the NP-hardness of the so-called
 {\em incomplete} MQC decision
problem, the less general {\em complete MQC decision problem} requires
$P$ to contain precisely one quartet topology per quartet out of $N$,
and is proven to be NP-hard as well in \cite{BJKLW99}.

\begin{theorem}
The MQTC decision problem is NP-hard.
\end{theorem}
\begin{proof}
By reduction from the MQC decision problem.
   For every MQC decision problem one can define a corresponding
MQTC decision problem that has the same solution: give the
quartet topologies in $P$ cost 0 and the ones in $Q - P$ cost 1.
Consider the MQTC decision problem: is there a tree $T$ with the
set $N$ labeling its leaves such that $C_T < {n \choose 4} -k$?
An alternative equivalent formulation is: is there a tree $T$ with the
set $N$ labeling its leaves such that
\[
S(T) > \frac{M- {n \choose 4}+k}{M-m} ?
\]
Note that every tree $T$ with the
set $N$ labeling its leaves has precisely one out of the three
quartet topologies of every of the ${n \choose 4}$ quartets
embedded in it. Therefore, the cost $C_T = {n \choose 4}-|P \bigcap Q_T |$.
If the answer to the above question is affirmative,
then the number of quartet topologies in $P$ that
are embedded in the tree exceeds $k$; if it is not then there
is no tree such that the number of quartet topologies in $P$
embedded in it exceeds $k$.
This way the MQC decision problem can be
reduced to the MQTC decision problem, which shows also the latter
to be NP-hard.
\end{proof}

Is it possible that the best $S(T)$ value is always one, that is,
there always exists a tree that embeds all quartets at minimum
cost quartet topologies?
Consider the case $n=|N|=4$. Since there is only one quartet,
we can set $T_0$ equal to the minimum cost quartet topology,
and have $S(T_0)=1$.
A priori we cannot exclude the possibility that
for every $N$ and $C$ there always is a tree $T_0$ with $S(T_0)=1$.
In that case, the MQTC Problem reduces to finding that $T_0$.
However, the situation turns out to be more complex. Note first
that the set of quartet topologies uniquely determines
a tree in ${\cal T}$, \cite{Bu71}.

\begin{lemma}
Let $T,T'$ be different labeled trees in ${\cal T}$ and let $Q_T,Q_{T'}$ be
the sets of embedded quartet topologies, respectively. Then,
$Q_T \neq Q_{T'}$.
\end{lemma}

A {\em complete set} of quartet topologies on $N$ is a set containing
precisely one quartet topology per quartet.
There are $3^{n \choose 4}$ such combinations, but only $2^{n \choose 2}$
labeled undirected graphs on $n$ nodes
(and therefore $|{\cal T}| \leq 2^{n \choose 2}$).
Hence, not every complete set of quartet topologies
corresponds to a tree in ${\cal T}$. This already suggests that we can
weight the quartet topologies in such a way that the full combination
of all quartet topologies at minimal costs does not correspond to
a tree in ${\cal T}$, and hence $S(T_0) < 1$ for $T_0 \in {\cal T}$
realizing the MQTC optimum. For an explicit example of this, we use
that a complete set corresponding to a tree in ${\cal T}$
 must satisfy certain transitivity properties,
 \cite{CS77,CS81}:

\begin{lemma}\label{cl1}
Let $T$ be a tree in the considered class with leaves $N$, $Q$ the set
of quartet topologies and
$Q_0 \subseteq Q$. Then
$Q_0$ uniquely determines $T$ if

(i) $Q_0$ contains precisely one quartet topology for every quartet,
and

(ii) For all $\{a,b,c,d,e\} \subseteq N$,
if $ab|bc, ab|de \in Q$ then $ab|ce \in Q$, as well as
if $ab|cd, bc|de \in Q$ then $ab|de \in Q$.
\end{lemma}

\begin{theorem}
There are $N$ (with $n=|N|=5$) and a cost function $C$ such that,
for every $T \in {\cal T}$,
$S(T)$ does not exceed $4/5$.
\end{theorem}
\begin{proof}
Consider $N=\{u,v,w,x,y\}$ and $C(uv|wx) = 1-\epsilon (\epsilon > 0),
C(uw|xv)= C(ux|vw)=0$,
$C(xy|uv)=C(wy|uv)=C(uy|wx)=C(vy|wx)=0$, and $C(ab|cd)=1$ for
all remaining quartet topologies $ab|cd \in Q$.
We see that $M= 5 - \epsilon$, $m=0$.
The tree $T_0 = (y,((u,v),(w,x)))$ has cost $C_{T_0}= 1-\epsilon$,
since it embeds quartet topologies $uw|xv, xy|uv, wy|uv, uy|wx, vy|wx$.
We show that $T_0$ achieves the MQTC optimum.
{\em Case 1:}
If a tree $T \neq T_0$ embeds $uv|wx$, then it
must by Item (i) of Lemma~\ref{cl1}
also embed a quartet topology
containing $y$ that has cost 1.

{\em Case 2:}
If a tree $T \neq T_0$ embeds $uw|xv$ and $xy|uv$, then it must by
Item (ii) of the Lemma~\ref{cl1}
also embed $uw|xy$,
and hence have cost $C_T \geq 1$. Similarly, all other
remaining cases of embedding a combination of a quartet
topology not containing $y$ of 0 cost with a quartet topology containing
$y$ of 0 cost in $T$, imply an embedded
quartet topology of cost 1 in $T$.
\end{proof}

Altogether,
the MQTC optimization problem is infeasible in practice, and natural data
can have an optimal $S(T)< 1$. In fact, it follows from the above
analysis that to determine $S(T)$ in general is NP-hard. In \cite{BJKLW99} a
polynomial time approximation scheme for complete MQC is exhibited,
a theoretical approximation scheme allowing the approximation
of the optimal solution up to arbitrary precision, with running time polynomial
in the inverse of that precision. We say ``theoretical'' since that algorithm
would run in something like $n^{19}$. For incomplete MQC it is
shown that even such a theoretical algorithm does not exist, unless
P=NP. Hence, computation of the MQTC optimum, and even its
 approximation with given precision, requires superpolynomial time
unless P=NP. Therefore, any practical approach to obtain or approximate the
MQTC optimum requires heuristics. 
%The most widely used method is Tree Puzzle
%\cite{SvH96} using consensus; there are many other
%methods like the geometric algorithim and dynamic
%programming in \cite{BCGOP98}, linear programming in
%\cite{WDGG05}, and quartet cleaning
%(The reduction also shows that the quartet problems reviewed in
%\cite{Ji01},
%are subsumed by our problem.)
%methods in \cite{Br00}.
%All of these techniques (and other previous methods not mentioned)
 %result in far too computationally
%intensive calculations;
%they run many months or years on moderate-sized problems
%of 30 objects.
%The method presented in this paper is a
%simple, feasible, heuristic based
%on randomization and hill-climbing that has been shown
%to handle routinely up to 60--80
%objects in a couple of hours \cite{CV04,We05}.

\section{New Heuristic}
Our algorithm is essentially randomized hill-climbing, using
parallellized Genetic Programming,  where
undirected trees evolve in a random walk 
driven by a prescribed fitness function. We are given a set $N$ of 
$n$ objects and a weighting function $W$. 

\begin{definition}
We define a {\em simple mutation} on a labeled undirected ternary tree
as one of three possible transformations:
\begin{enumerate}
\item A {\em leaf swap}, which consists of randomly choosing two leaf nodes
and swapping them.
\item A {\em subtree swap}, which consists of randomly choosing two internal
nodes and swapping the subtrees rooted at those nodes.
\item A {\em subtree transfer}, whereby a randomly chosen subtree (possibly a leaf) is detached and reattached in another place, maintaining arity invariants.
\end{enumerate}
\end{definition}
Each of these simple mutations keeps the
number of leaf nodes and internal nodes in the tree invariant;
only the structure and placements
change.
\begin{definition}
A {\em $k$-mutation} is a sequence of $k$
simple mutations.
Thus, a simple mutation is a 1-mutation.
\end{definition}

\subsection{Algorithm}

{\bf Step 1:} First, a random tree $T \in {\cal T}$ with $2n-2$ nodes
is created, consisting of $n$ leaf nodes (with 1 connecting edge) labeled
with the names of the data items, and $n-2$ non-leaf or {\em internal} nodes
labeled with the lowercase letter ``k'' followed by a
unique integer identifier.  Each internal node has exactly
three connecting edges.

{\bf Step 2:} For this
tree $T$, we calculate the total cost of all embedded quartet topologies,
compute  $S(T)$.

{\em Comment:} A tree is consistent with precisely
$\frac{1}{3}$ of all quartet topologies, one for every quartet.
A random tree is likely to be consistent with about $\frac{1}{3}$ of the best
quartet topologies---but because of dependencies this figure is
not precise.

{\bf Step 3:}
The {\em currently best known tree} variable $T_0$ is set
to $T$:   $T_0 \leftarrow T$.

{\em Comment:} This $T_0$  is used as
the basis for further searching.

{\bf Step 4:}
Pick a number $k$ with
probability $p(k) = c /(k (\log k)^2)$ where
$1/c = \sum_{k=1}^{\infty} 1/(k (\log k)^2)$.

{\em Comment:}
This number $k$ is the number of simple mutations that we will perform
in the next $k$-mutation. The probability distribution $p(k)$
is easily generated by running a random fair bit generator and
set $k$ to the length of the first self-delimiting sequence generated.
That is, if $x=x_1 \ldots x_k \in \{0,1\}^k$ ($|x|=k \geq 1$),
then $\bar{x}= 1^{k-1} 0 x$, $x'= \overline{|x|}x$, and
$x'' = \overline{|x'|}x'$. Thus, the length $|x''|=k + \log k + 2 \log \log k $.
The probability
of generating $x''$ corresponding to a given $x$ of length $k$
by fair coin flips is $ 2^{-|x''|}= 2^{-k- \log k - 2 \log \log k}
= 2^{-k}/(k (\log k)^2)$. The probability of generating $x''$
corresponding to {\em some} $x$ of length $k$ is $2^k$ times as
large, that is, $1/(k (\log k)^2)$. In practice, we used a ``shifted'' fat tail
distribution $1/((k+2) (\log k+2)^2)$

{\bf Step 5:}
Compose a $k$-mutation by,
for each such simple mutation, choosing one of
the three types listed above with equal probability.  For each of
these simple mutations, we uniformly at random select
 leaves or internal nodes, as appropriate.

{\em Comment:} Notice
that trees which are close to the original tree (in terms of number of
simple mutation steps in between) are examined often, while trees that are
far away from the original tree will eventually be examined, but not very
frequently.

{\bf Step 6:}
In order to search for a better tree,
we simply
apply the $k$-mutation constructed in {\bf Step 5}
 on $T_0$ to obtain $T'$, and then
calculate $S(T')$.  If $S(T') \geq S(T_0)$, then replace the current
candidate in $T_0$ by $T$ (as the new best tree):
$T_0 \leftarrow T$.

{\bf Step 7:}
If $S(T_0) =1$ or a {\bf termination condition} to be discussed below holds,
then output the tree in $T_0$ as the best tree and halt.
Otherwise, go to {\bf Step 4}.

\begin{remark}
\rm
We have chosen $p(k)$ to be a ``fat-tail'' distribution,
with the fattest tail possible,
to concentrate maximal probability also on the larger values of $k$.
That way, the likelihood of getting trapped in local minima is minimized.
In contrast, if one would choose an exponential scheme, like
$q(k)=c e^{-k}$, then the larger values of $k$ would arise so scarcely
that practically speaking the distinction between being absolutely trapped in
a local optimum, and the very low escape probability, would be
insignificant. Considering positive-valued probability mass
functions $q: {\cal N} \rightarrow (0,1]$, with ${\cal N}$
the natural numbers, as we do here, we note that such a function
(i) $\lim_{k \rightarrow \infty} q(k) =0$,
 and (ii) $\sum_{k=1}^{\infty} q(k) =1$.
Thus, every function of the natural numbers
that has stricly positive values and converges can be normalized to such
a probability mass function. For smooth analytic functions that can be expressed
a series of fractional powers and logarithms, the borderline between
converging and diverging is as follows: $\sum 1/k, \sum 1/(k \log k)$,
$\sum 1/(k \log k \log \log k)$ and so on diverge, while
$\sum 1/k^2, \sum 1/(k (\log k)^2)$,$\sum 1/(k \log k (\log \log k)^2)$
 and so on
converge. Therefore,
the maximal fat tail of a ``smooth'' function $f(x)$
with $\sum f(x) < \infty$  arises for functions at the edge of the convergence
family. The distribution $p(k)= c /(k (\log k)^2)$ is as close to the edge
as is reasonable, and because the used coding $x \rightarrow x''$
is a prefix code we have $\sum 1/(k (\log k)^2) \leq 1$ by the Kraft Inequality
(see for ecample \cite{LiVi97}) and therefore $c \geq 1$. Let us see what this
means for our algorithm using the choosen distribution $p(k)$.
For $N=64$, say, we can change any tree in ${\cal T}$ to any other tree in
${\cal T}$ with a 64-mutation. The probability of such a complex mutation
occurring is quite large with such a fat tail: $1/(64 \cdot 6^2) = 1/2304$, that is,
more than 40 times in 100,000 generations. If we can get out of a local
minimum with already a 32-mutation, then this occurs with probability
at least $1/800$, so 125 times, and with a 16-mutation with probability
at least $1/196$, so 510 times.
\end{remark}

\subsection{Performance}
The main problem with hill-climbing algorithms is that they can get stuck
in a local optimum. However, by randomly selecting a sequence of simple
mutations, longer sequences with decreasing probability, we essentially
run a Metropolis Monte Carlo algorithm \cite{MRRTT53}, reminiscent
of simulated annealing \cite{KGV83} at random temperatures.
Since there is a nonzero probability for every tree in ${\cal T}$
being transformed into every other tree in ${\cal T}$, there is zero
probability that we get trapped forever
in a local optimum that is not a global optimum.
That is, trivially:
\begin{lemma}
(i) The algorithm approximates the MQTC optimal solution monotonically
in each run.

(ii) The algorithm without termination condition
solves the MQTC optimization problem eventually with probability 1
(but we do not in general know when the optimum has been
reached in a particular run).
\end{lemma}

\begin{figure}[htb]
\begin{center}
\epsfig{file=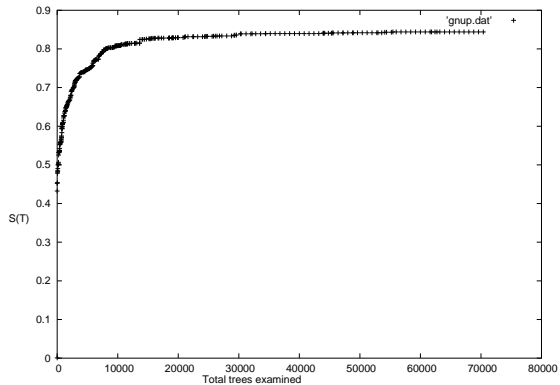,width=2in,angle=270}
\end{center}
\caption{Progress of a 60-item data set experiment over time}\label{figprogress}
\end{figure}

The main question therefore is the convergence speed of the algorithm
on natural data, and a termination criterion to terminate the
algorithm when we have an acceptable approximation. From the
impossibility result in \cite{BJKLW99} we know that there is no
polynomial approximation scheme for MQTC optimization, and
whether our scheme is expected polynomial time seems to require
proving that the involved Metropolis chain is rapidly mixing
\cite{Vi00}, a notoriously hard and generally unsolved problem.
In practice, in our experiments there is
unanymous evidence that for the natural data and the weighting
function we have used, convergence is always fast. We have to
determine the cost of ${n \choose 4}$ quartets to determine
each $S(T)$ value. Hence the algorithm runs in time at least that
much. In experiments we found that for the same data set different
runs consistently showed the same behavior, for example
Figure~\ref{figprogress} for a 60-object computation. There
the $S(T)$ value leveled off at about 70,000 examined trees,
and the termination condition was ``no improvement in 5,000 trees.''
Different random runs of the algorithm nearly always gave the same behavior,
returning a tree with the same $S(T)$ value, albeit a different tree in
most cases with here $S(T) \approx 0.865$, a relatively low value.
That is, since there are many ways to find a tree of optimal
$S(T)$ value, and apparently the algorithm never got trapped
in a lower local optimum. For problems with high $S(T)$ value, as we
see later, the algorithm consistently returned the same
tree. This situation is perhaps similar to the behavior of the
Simplex method in linear programming, that can be shown to run
in exponential time on a badly choosen problem instance, but in practice
on natural problems consistently runs in linear time.

Note that if a tree is ever found such that $S(T) = 1$, then we can stop
because we can be certain that this tree is optimal, as no tree could
have a lower cost.  In fact, this perfect tree result is achieved in our
artificial tree reconstruction experiment (Section~\ref{sect.artificial})
reliably in a few minutes.  For real-world data, $S(T)$ reaches
a maximum somewhat
less than $1$, presumably reflecting distortion of the information
in the distance matrix
data by the best possible tree representation, as noted above,
or indicating getting stuck in a local optimum or a search space too large
to find the global optimum.
On many typical problems of up to 40 objects this tree-search gives a tree
with $S(T) \geq 0.9$ within half an hour.  For large numbers of objects,
tree scoring itself can be slow: as this takes order $n^4$ computation steps.
Current single computers can score a tree of this size in about a minute.
Additionally, the space of trees is large, so the algorithm may slow down
substantially.  For larger experiments, we used the C program called
partree (part of the CompLearn package \cite{Ci03}) 
with MPI (Message Passing Interface, a common
standard used on massively parallel computers) on a cluster of workstations in
parallel to find trees more rapidly.  We can consider the graph
mapping the achieved $S(T)$ score as a function
of the number of trees examined.  Progress
occurs typically in a sigmoidal fashion towards a maximal value $\leq 1$,
Figure~\ref{figprogress}.

\subsection{Termination Condition}
The {\em termination condition} is of two types and which type
is used determines the number of objects we can handle.

{\em Simple termination condition:} We simply run the
algorithm until it seems
no better trees are being found in a reasonable amount of time.
Here we typically terminate if no improvement in $S(T)$ value is
achieved within 100,000 examined trees. This criterion is simple
enough to enable us to hierarchically cluster data sets up to 80
objects in a few hours. This is way above the 15--30 objects in
the previous exact (non-incremental) methods (see Introduction).

{\em Agreement termination condition:} In this more sophisticated method we
select a number $2 \leq r \leq 6$ of runs, and we run $r$ invocations
of the algorithm in parallel. Each time an $S(T)$ value in run $i=1, \ldots, r$
is increased in this process it is compared with the $S(T)$ values
in all the other runs. If they are all equal, then the candidate trees
of the runs are compared. This can be done by simply comparing the
ordered lists of embedded quartet topologies, in some standard order,
since the  set of embedded quartet topologies uniquely
determines the quartet tree by \cite{Bu71}. If the $r$ candidate trees
are identical, then terminate with this quartet tree as output, otherwise
continue the algorithm.

This termination condition takes (for the same number of steps per run)
about $r$ times as long as the simple termination condition.
But the termination
condition is much more rigorous, provided we choose $r$
appropriate to the number $n$
of objects being clustered.
 Since all the runs are randomized independently at startup,
it seems very unlikely that with natural data
all of them get stuck in the same local
optimum with the same quartet tree instance,
provided the number $n$ of objects being clustered is not
too small. For $n = 5$ and the number of invocations $r=2$,
there is a reasonable probability that the two different
runs by chance hit the same tree in the same step. This phenomenon
leads us to require more than two successive runs with exact agreement before
we may reach a final answer for small $n$.  In the case of $4\le n \le 5$, we
require 6 dovetailed runs to agree precisely before termination.  For $6 \le n
\le 9$, $r = 5$. For $10 \le n \le 15$, $r = 4$.  For $16 \le n \le 17$,
$r = 3$.  For all other $n \ge 18$, $r = 2$.  This yields a reasonable tradeoff
between speed and accuracy.

It is clear that there is only one tree with $S(T)=1$ (if that is possible for
the data), and random trees (the majority of all possible quartet trees) have
$S(T) \approx 1/3$ (above).  This gives evidence that the number of quartet
trees with large $S(T)$ values is much smaller than the number of trees with
small $S(T)$ values.  It is furthermore evident that the precise relation
depends on the data set involved, and hence cannot be expressed by a general
formula without further assumptions on the data. However, we can
safely state that small data sets, of say $\leq 15$ objects,
that in our experience often lead to $S(T)$ values close to 1 have
very few quartet trees realizing the optimal $S(T)$ value. On the other
hand, those large sets of 60 or more objects that contain some inconsistency
and thus lead to a low final $S(T)$ value also tend to exhibit more variation
as one might expect.
This suggests that in the agreement
termination method each run will get stuck in a different quartet
tree of a similar $S(T)$ value, so termination with the same tree
is not possible. Experiments show that with the rigorous agreement termination
we can handle sets of up to 40 objects,
%
%Both methods improve all existing quartet methods in at least two ways:
%
%(i) Existing quartet methods are all incremental: first two quartets
%are merged to a tree based on some criterion, then a next quartet is
%added, and so on. While the choice of the first two quartets optimizes
%the used criterion, this is not so for the next incremental steps.
%Since the optimal tree is optimal according to a global cost, this may
%mean that no subtree is optimal according to that cost. Thus, a wrong
%initial choice in an incremental method cannot be undone at a later stage.
%In contrast, in our approach the total tree is modified to optimize the
%global cost, and hence monotonically approximates the global optimum.
%
%(ii) In none of the previous methods quartet trees of more than 20
%objects can be directly handled, while our method with the agreement
%termination handles up to 40 objects, 
and with the simple termination
up to at least 80 objects on a single computer or 100-200 objects using
a cluster of computers in parallel.

\subsection{Tree Building Statistics}
\begin{figure}[htb]
\begin{center}
\epsfig{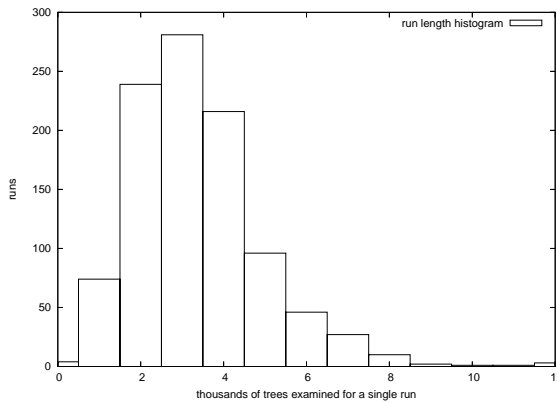}
\end{center}
\caption{Histogram of run-time number of trees examined before termination.}\label{fig.plot}
\end{figure}

We used the CompLearn package, \cite{Ci03},
to analyze a ``10-mammals'' example with {\em zlib} compression
yielding a  $10 \times 10$ distance matrix, similar to the examples in 
Section~\ref{sect.nat}.
The algorithm starts with four randomly initialized trees.
It tries to improve each one randomly and finishes when they match.
Thus, every run produces an output tree, a maximum score associated
with this tree, and has examined some total number of trees,
$T$, before it finished.
%We may imagine that $T$ comes from a distribution,
%getting one sample per run of the tree reconstruction program for
%a given distance matrix.
Figure~\ref{fig.plot}
shows a graph displaying a histogram
of $T$ over one thousand runs of the distance matrix.  The $x$-axis
represents a number of trees examined in a single run of the program, measured
in thousands of trees and binned in 1000-wide histogram bars.  The maximum
number is about 12000 trees examined.  
The graph suggests a Poisson distribution.
 About $2/3$rd of the trials take less than 4000 trees.  In the thousand trials
above, 994 ended with the optimal $ S(T) = 0.999514 $.  The remaining six runs
returned 5 cases of the second-highest score, $  S(T) = 0.995198 $ and one case
of $ S(T) = 0.992222 $.  It is important to realize that outcome stability is
dependent on input matrix particulars.

\begin{figure}[htb]
\begin{center}
\epsfig{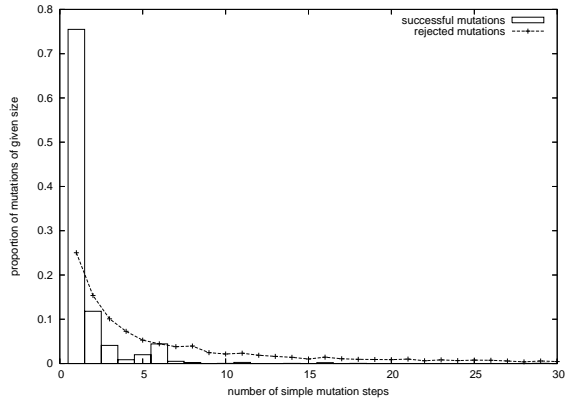}
\end{center}
\caption{Histogram comparing distributions of $k$-mutations per run.}\label{fig.mutplot}
\end{figure}

Another interesting distribution is the mutation stepsize.
Recall that the mutation length is drawn from a shifted fat-tail distribution.
But if we restrict our attention to just the mutations
that improve the $S(T)$ value, then we may examine these statistics
to look for evidence of a modification to this distribution due to,
for example, the presence of very many isolated areas that have only
long-distance ways to escape.  Figure~\ref{fig.mutplot}
 shows the histogram
of successful mutation lengths (that is, number of simple mutations
composing a single “kept” complex mutation) and rejected lengths (both
normalized) which shows that this is not the case.
Here the $x$-axis is the number of mutation steps and the $y$-axis is
the normalized proportion of times that step size occured.
This gives good empirical evidence that in this case,
at least, we have a relatively easy search space, without large gaps.

\subsection{Controlled Experiments}\label{sect.artificial}

With natural data sets, say music data, one may have the preconception
(or prejudice) that  music by Bach should be clustered together,
music by Chopin should be clustered together, and so should music by
rock stars. However, the preprocessed music files of a piece by Bach and
a piece by Chopin, or the Beatles, may resemble one another
more than two different
pieces by Bach---by accident or indeed by design and copying. Thus, natural
data sets may have ambiguous, conflicting, or counterintuitive
outcomes. In other words, the experiments on natural data sets have
the drawback of not having an objective clear ``correct'' answer that can
function as a benchmark for assessing our experimental outcomes,
but only intuitive or traditional preconceptions.
We discuss three experiments that show that our
program indeed does what it is supposed to do---at least in
artificial situations where we know in advance what the correct answer is.

\section{Quartet Topology Costs based on Distance Matrix}
Given a distance matrix, with entries giving the pairwise
distances between the objects, we want to hierarchically cluster them
by representing the objects as leaves of a ternary tree
representing the distances in the matrix as faithfully as possible.
It is important that we do not assume that there is a true tree;
rather, we want to model the data as well as possible.
The cost of a quartet topology is defined as the sum
of the distances between each pair of neighbors; that
is, $C_{uv|wx} = d(u,v) + d(w,x)$. This seems most natural given
a distance matrix.

\subsection{Distance Measure Used}
Recall that the problem of clustering data
consists of two parts: (i) extracting a distance matrix
from the data, and (ii) constructing a tree
from the distance matrix using our novel quartet-based heuristic.
To check the new quartet tree method in action we use a new
compression-based distance, called \NCD.
This metric distance
was co-developed by us in \cite{LBCKKZ01,Li01,Li03}, as a normalized
version of the ``information metric'' of \cite{LiVi97,BGLVZ98}.
Roughly speaking, two objects are deemed close if
we can significantly ``compress'' one given the information
in the other, the idea being that if two pieces are more similar,
then we can more succinctly describe one given the other.
The mathematics used is based on Kolmogorov complexity theory \cite{LiVi97}.
In \cite{Li03} we defined a
new class of (possibly non-metric) distances, taking values in $[0,1]$ and
appropriate for measuring effective
similarity relations between sequences, say one type of similarity
per distance, and {\em vice versa}. It was shown that an appropriately
``normalized'' information metric
minorizes every distance
in the class.
It discovers all effective similarities in the sense that if two
objects are close according to some effective similarity, then
they are also close according to the normalized information distance.
Put differently, the normalized information distance represents
similarity according to the dominating shared feature between
the two objects being compared.
In comparisons of more than two objects,
different pairs may have different dominating features.
The normalized information distance is a metric
and takes values in $[0,1]$;
hence it may be called {\em ``the'' similarity metric}.
To apply this ideal precise mathematical theory in real life,
we have to replace the use of  the noncomputable
Kolmogorov complexity by an approximation
using a standard real-world compressor, resulting in the
\NCD, see \cite{CV04}. This has been used
 in
the first completely automatic construction
of the phylogeny tree based on whole mitochondrial genomes,
\cite{LBCKKZ01,Li01,Li03},
a completely automatic construction of a language tree for over 50
Euro-Asian languages \cite{Li03},
detects plagiarism in student programming assignments
\cite{SID}, gives phylogeny of chain letters \cite{BLM03}, and clusters
music \cite{CVW03},
Analyzing network traffic and worms using compression \cite{We05},
and many more topics \cite{CV04}.
The method turns out to be robust under change of the underlying
compressor-types: statistical (PPMZ), Lempel-Ziv based  dictionary (gzip),
block based (bzip2), or special purpose (Gencompress).

\subsection{CompLearn Toolkit}
Oblivious to the problem area concerned, simply using the distances
according to the \NCD above,
the method described in this paper fully automatically
classifies the objects concerned.
The method has been released in the public domain as open-source software:
The CompLearn Toolkit \cite{Ci03} is a suite
of simple utilities that one can use to apply compression
techniques to the process of discovering and learning patterns
in completely different domains, and hierarchically cluster them
using the new quartet method described in this paper.
In fact, this method is so general that it requires
no background knowledge about any particular
subject area. There are no domain-specific parameters to set,
and only a handful of general settings.

\begin{figure}[htb]
\begin{center}
\epsfig{file=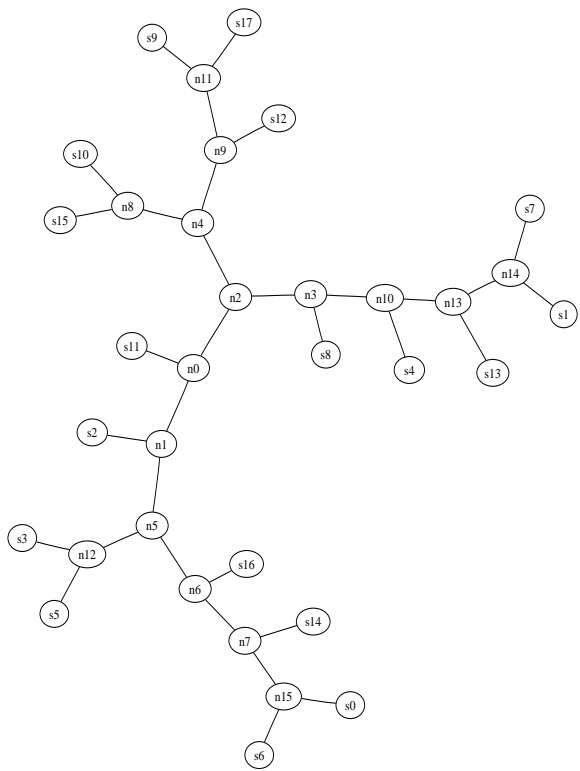,width=3in,height=4in}
\end{center}
\caption{The randomly generated tree that our algorithm reconstructed. $S(T)=1$.}\label{figarttreereal}
\end{figure}

\subsection{Testing The Quartet-Based Tree Construction}
We first test whether the quartet-based tree construction
heuristic is trustworthy:
We generated a ternary tree $T$ with 18 leaves, using the pseudo-random
number generator ``rand'' of the Ruby programming language,
and derived
a metric from it by defining the distance between
two nodes as follows:
Given the length of the path from $a$ to $b$, in an integer number of
edges, as $L(a,b)$, let
\[d(a,b) = { {L(a,b)+1} \over 18},
\]
  except when
$a = b$, in which case $d(a,b) = 0$.  It is easy to verify that this
simple formula always gives a number between 0 and 1, and is monotonic
with path length.
Given only the $18\times 18$ matrix of these normalized distances,
our quartet method exactly reconstructed  the original tree
$T$ represented in
Figure~\ref{figarttreereal},
with $S(T)=1$.
%TODO: Rudi, Paul wants the distance matrix included here as well

\begin{figure}[htb]
\begin{center}
\epsfig{file=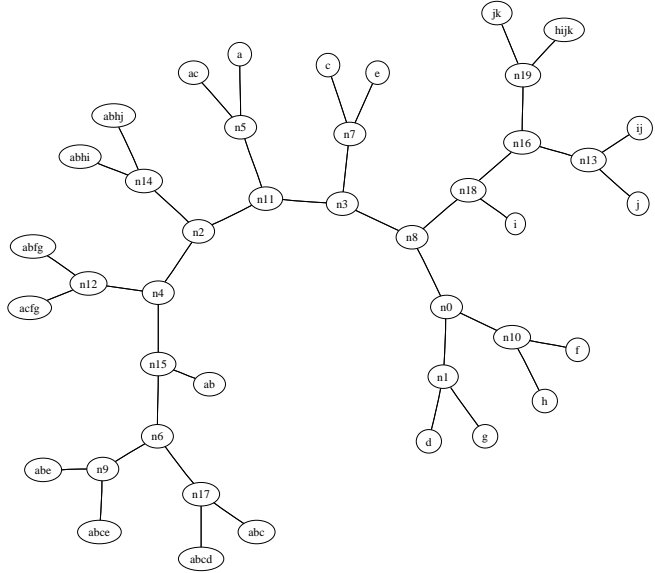,height=3in}
\end{center}
\caption{Classification of artificial files with repeated 1-kilobyte tags.
 Not all possiblities
are included; for example, file ``$b$'' is missing.
$S(T)=0.905$. }\label{figtaggedfiles}
\end{figure}
\begin{figure}[htb]
\begin{center}
\epsfig{file=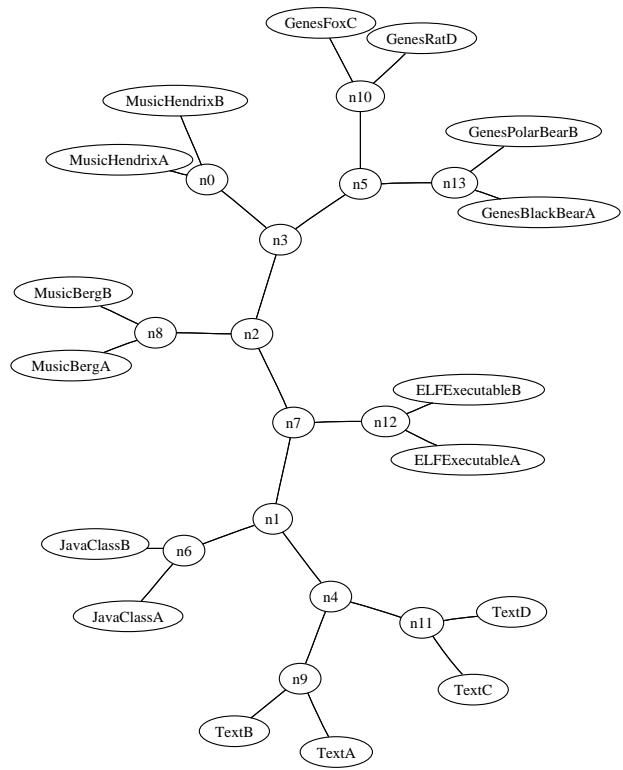,height=4in}
\end{center}
\caption{Classification of different file types.
Tree agrees
exceptionally well with
\NCD distance matrix: $S(T)=0.984$.}\label{figfiletypes}
\end{figure}

\section{Testing on artificial data}
Given that the tree reconstruction method is accurate
on clean consistent data, we tried whether the full procedure
works in an acceptable manner when we know what the outcome should
be like. We used the ``rand'' pseudo-random number generator
from the C programming language standard library under Linux.
We randomly generated 11 separate 1-kilobyte blocks of data where
each byte was equally probable and called these {\em tags}.  Each tag
was associated with a different lowercase letter of the alphabet.  Next,
we generated 22 files of 80 kilobyte each,
by starting with a block of purely random
bytes and applying one, two, three, or four different tags on it.
Applying a tag consists of ten repetitions of picking a random location
in the 80-kilobyte file, and overwriting that location with the globally
consistent tag that is indicated.  So, for instance, to create the file
referred to in the diagram by ``a,'' we start with 80 kilobytes of random data,
then pick ten places to copy over this random data with the arbitrary
1-kilobyte sequence identified as tag {\em a}.
Similarly, to create file ``ab,''
we start with 80 kilobytes of random data, then pick ten places to put
copies of tag {\em a}, then pick ten more places to put copies of tag {\em b} (perhaps
overwriting some of the {\em a} tags).  Because we never use more than four
different tags, and therefore never place more than 40 copies of tags, we
can expect that at least half of the data in each file is random and
uncorrelated with the rest of the files.  The rest of the file is
correlated with other files that also contain tags in common; the more
tags in common, the more related the files are.
The compressor used to compute the \NCD matrix was bzip2.
The resulting tree
 is given in Figure~\ref{figtaggedfiles}; it can be
seen that
the clustering has  occured exactly as we would expect.
The $S(T)$ score is 0.905.

\section{Testing On Heterogenous Natural Data}
We test gross classification of files
based on heterogenous data of markedly different file types:
(i) Four mitochondrial gene sequences, from a black bear, polar bear,
fox, and rat obtained from the GenBank Database on the world-wide web;
(ii) Four excerpts from the novel { \em The Zeppelin's Passenger} by
E.~Phillips Oppenheim, obtained from the Project Gutenberg Edition
on the World-Wide web;
(iii) Four MIDI files without further processing; two from Jimi Hendrix and
two movements from Debussy's Suite Bergamasque, downloaded from various
repositories on the
world-wide web;
(iv) Two Linux x86 ELF executables (the {\em cp} and {\em rm} commands),
copied directly from the RedHat 9.0 Linux distribution; and
(v)  Two compiled Java class files, generated by ourselves.
The compressor used to compute the \NCD matrix was bzip2.
As expected, the program correctly classifies each of the different types
of files together with like near like. The result is reported
in Figure~\ref{figfiletypes} with $S(T)$ equal to the very high
confidence value 0.984.
This experiment shows the power and universality of the method:
no features of any specific domain of application are used.
We believe that there is no other method known that can cluster
data that is so heterogenous this reliably. This is borne out by the
massive experiments with the method in \cite{Ke04}.

\section{Testing on Natural Data}
\label{sect.nat}
Like most hierarchical clustering methods for natural data,
the quartet tree method has been developed in the biological setting
to determine phylogeny trees from genomic data. In that setting, the
data are (parts of) genomes of currently existing species,
and the purpose is to reconstruct the evolutionary tree that led
to those species. Thus, the species are labels of the leaves,
and the tree is traditionaly binary branching with each branching
representing a split in lineages. The internal nodes and the root
of the tree correspond with extinct species (possibly a still
existing species in a leaf directly connected to the internal node).
The case is roughly similar for the language tree reconstruction
mentioned in the Introduction. The root of the tree is commonly
determined by adding an object that is known to be less related
to all other objects than the original objects are with respect to
each other. Where the unrelated object joins the tree is where
we put the root.
In these settings, the direction from the root to the leaves represents
an evolution in time, and the assumption is that there is a true
tree we have to discover.
However, we can also use the method for hierarchical clustering,
resulting an unrooted ternary tree, and the assumption is not that there
is a true tree we must discover. To the contrary, there is no true
tree, but all we want is to model the similarity relations between
the objects as well as possible, given the distance matrix.
The interpretation is that objects in a given subtree are pairwise
closer (more similar) to each other than any of those objects
is with respect to any object in a disjoint subtree.

\subsection{Analyzing the SARS and H5N1 Virus Genomes}
\begin{figure}
\hfill\ \psfig{figure=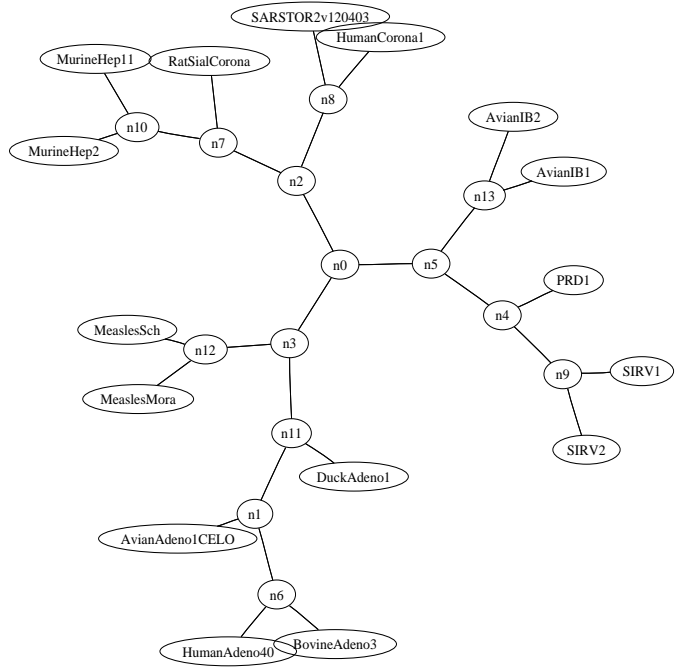,width=3.5in,height=3.5in} \hfill\
\caption{SARS virus among other virii. Legend:
AvianAdeno1CELO.inp:  Fowl adenovirus 1;
AvianIB1.inp:  Avian infectious bronchitis virus (strain Beaudette US);
AvianIB2.inp:  Avian infectious bronchitis virus (strain Beaudette CK);
BovineAdeno3.inp:  Bovine adenovirus 3;
DuckAdeno1.inp:  Duck adenovirus 1;
HumanAdeno40.inp:  Human adenovirus type 40;
HumanCorona1.inp:  Human coronavirus 229E;
MeaslesMora.inp:  Measles virus strain Moraten;
MeaslesSch.inp:  Measles virus strain Schwarz;
MurineHep11.inp:  Murine hepatitis virus strain ML-11;
MurineHep2.inp:  Murine hepatitis virus strain 2;
PRD1.inp:  Enterobacteria phage PRD1;
RatSialCorona.inp:  Rat sialodacryoadenitis coronavirus;
SARS.inp: SARS TOR2v120403;
SIRV1.inp:  Sulfolobus virus SIRV-1;
SIRV2.inp:  Sulfolobus virus SIRV-2.
$S(T)=0.988$.
}\label{fig.sars}
\end{figure}

As an application of our methods we clustered the SARS virus after its
sequenced genome was made publicly available,
 in relation to potential similar virii.
The 15 virus genomes were downloaded from The Universal Virus
Database of the International Committee on Taxonomy of Viruses,
available on the world-wide web.
The SARS virus was downloaded from
Canada's Michael Smith Genome Sciences Centre which
had the first public SARS Coronovirus draft whole genome
assembly available for download (SARS TOR2 draft genome assembly 120403).
The \NCD distance matrix was computed using the compressor bzip2.
The relations in Figure~\ref{fig.sars} are very similar to the
definitive tree based on medical-macrobio-genomics analysis,
appearing later in the New England Journal of Medicine,
\cite{SA03}.
We depicted the figure in the ternary tree style, rather than the
genomics-dendrogram style, since the former is more precise
for visual inspection of proximity relations.

More recently, we downloaded 100 different H5N1 sample genomes from the NCBI/NIH
database online.  We simply concatenated all data together directly, ignoring
problems of data cleanup and duplication.  We were warned in advance that
certain coding regions in the viral genome were sometimes listed twice and
also many sequences are incomplete or missing certain proteins.  In this case
we sought to test the robustness at the high end and at the same time verify,
contextualize, and expand on the many claims of genetic similarity and
diversity in the virology community.  We used the CompLearn package, \cite{Ci03},
with the {\em ppmd} compressor for
this experiment and performed no alignment step whatsoever.  We used order
15 with 250 megabytes memory maximum.

\begin{figure*}
\begin{center}
\epsfig{figure=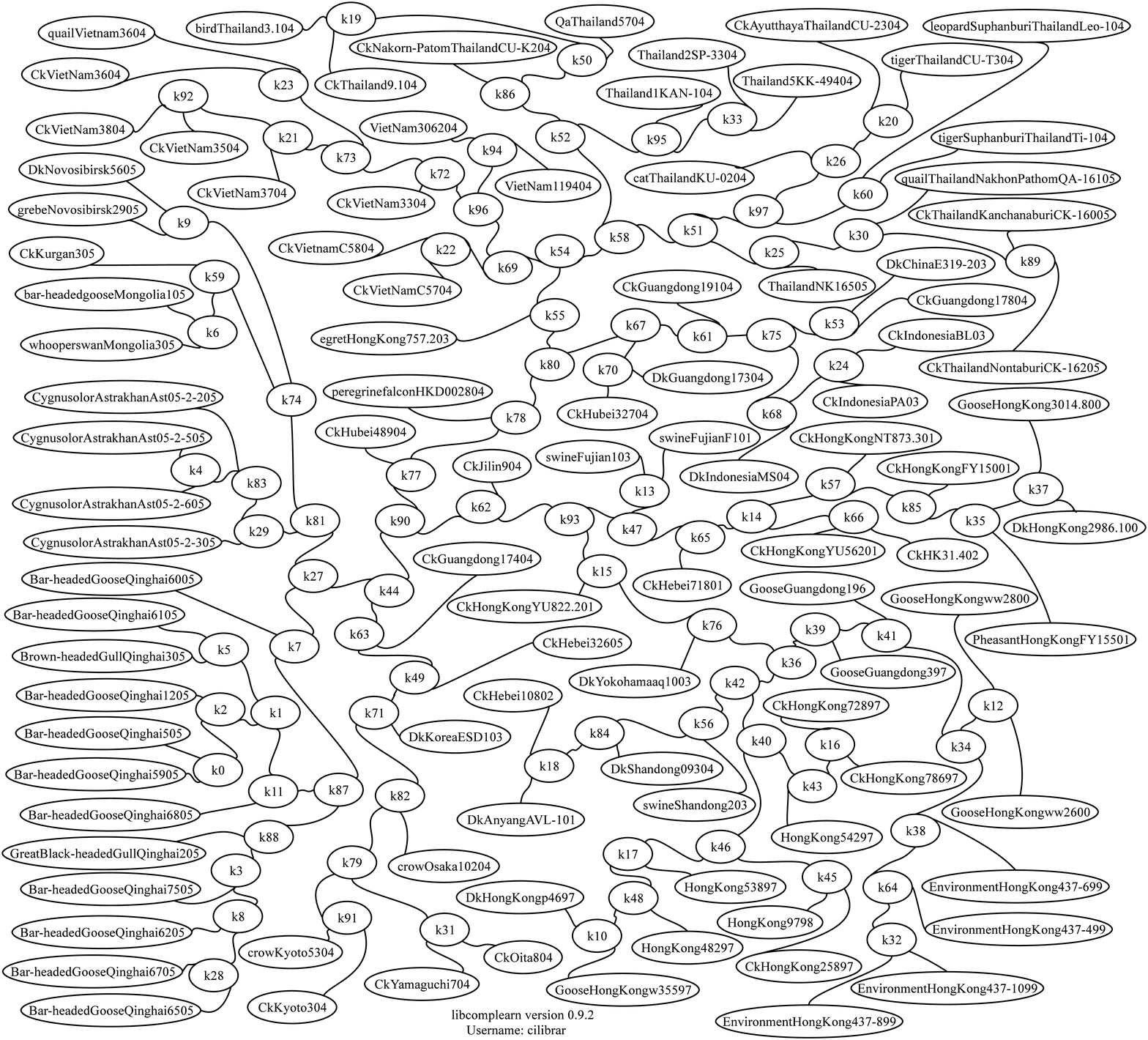,width=6.5in,height=9in}
\caption{One hundred H5N1 (bird flu) sample genomes, S(T) = 0.980221 }
\label{fig.big100}
\end{center}
\end{figure*}

We have abbreviated Ck for Chicken and Dk for duck.  Samples are named
with species, location, sequence number, followed by the year double digits at the
end.  Naming is not 100\% consistent.  We can see the
following features in Figure~\ref{fig.big100}, that are possibly significant:

First, there is near-perfect temporal separation by branch and year, going all
the way back to HongKong and GuangDong in 1997.
Next, there is near-perfect regional separation with clear dilineation of Japan
and the crucial Qinghai, Astrakhan, Mongolia, and Novosibirsk, as well as
near-perfect separation of Vietnam and Thailand. The placement
CkVietnamC5804 and Vietnam306204 is interesting in that they are both
near Thailand branches and suggest that they may be for example the
migratory bird links that have been hypothesized or some other genetic
intermediate.
There is also throughout the tree substantial agreement with (and independent
verification of) independent experts like Dr.~Henry L. Niman~\cite{Ni06} on every specific point regarding genetic similarity.  The technique provides here
an easy verification procedure without much work.

\subsection{Music}
\begin{figure}
\begin{center}
\epsfig{file=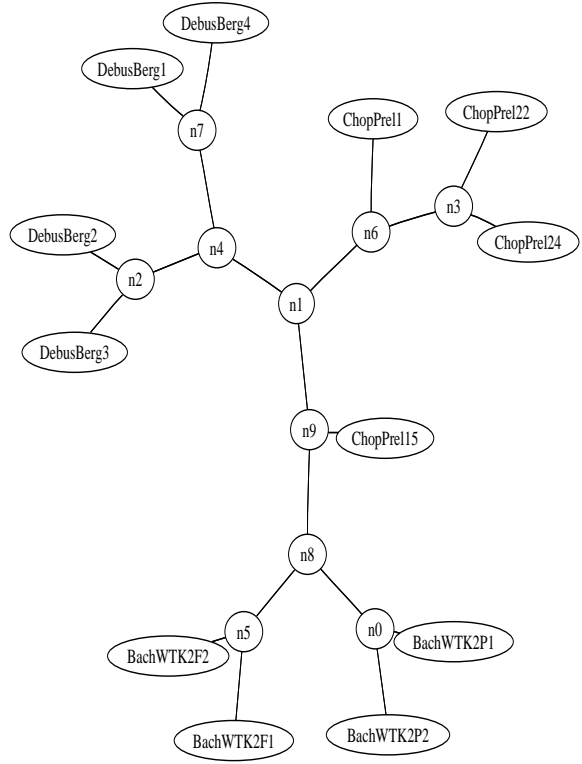,width=3in,height=4in}
\end{center}
\caption{Output for the 12-piece set.
Legend: J.S. Bach [Wohltemperierte Klavier II: Preludes and Fugues 1,2---
BachWTK2\{F,P\}\{1,2\}]; Chopin [Pr\'eludes op.~28: 1, 15, 22, 24
---ChopPrel\{1,15,22,24\}];
Debussy [Suite Bergamasque, 4 movements---DebusBerg\{1,2,3,4\}].
$S(T)=0.968$.}\label{figsmallset}
\end{figure}
The amount of digitized music available on the internet has grown
dramatically in recent years, both in the public domain
and on commercial sites. Napster and its clones are prime examples.
Websites offering musical content in some form or other
(MP3, MIDI, \ldots) need a way to organize their wealth of material;
they need to somehow classify their files according to
musical genres and subgenres, putting similar pieces together.
The purpose of such organization is to enable users
to navigate to pieces of music they already know and like,
but also to give them advice and recommendations
(``If you like this, you might also like\ldots'').
Currently, such organization is mostly done manually by humans,
but some recent research has been looking into the possibilities
of automating music classification. In \cite{CVW03} we cluster
music using the CompLearn Toolkit \cite{Ci03}.
One example is a small set of classical piano sonatas,
consisting of the 4 movements from Debussy's ``Suite Bergamasque,''
4 movements of book 2 of Bach's ``Wohltemperierte Klavier,''
and 4 preludes from
Chopin's ``Opus~28.'' As one can see in Figure~\ref{figsmallset},
our program does a pretty good job at clustering these pieces.
The $S(T)$ score is also high: 0.968.
The 4 Debussy movements form one cluster, as do the 4 Bach pieces.
The only imperfection in the tree, judged by what one would
intuitively expect, is that Chopin's Pr\'elude no.~15 lies a bit closer
to Bach than to the other 3 Chopin pieces.
This Pr\'elude no~15, in fact, consistently forms an odd-one-out
in our other experiments as well. This is an example of pure data mining,
since there is some musical truth
to this, as no.~15 is perceived as by far the most eccentric
among the 24 Pr\'eludes of Chopin's opus~28.

\subsection{Mammalian Evolution}
\begin{figure}
\hfill\ \psfig{figure=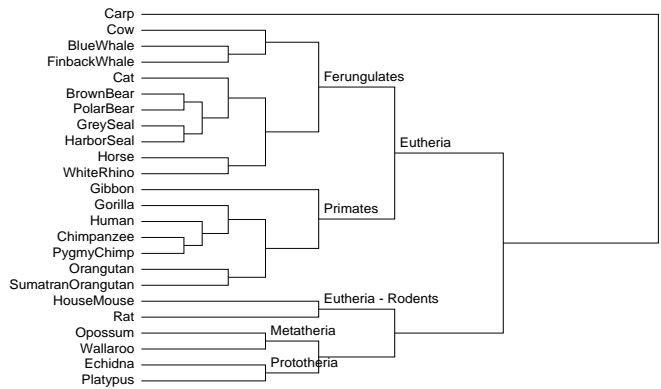,width=3.5in,height=2in} \hfill\
\caption{The evolutionary tree built from complete mammalian mtDNA
sequences of 24 species, using the \NCD matrix of Figure~\ref{fig.distmatr}.
We have redrawn the tree from our output to agree
better with the customary phylogeny tree format. The tree agrees exceptionally
well with the \NCD distance matrix: $S(T)=0.996$.
}
\label{tree-mammal}
\end{figure}
\begin{figure}
\hfill\ \psfig{figure=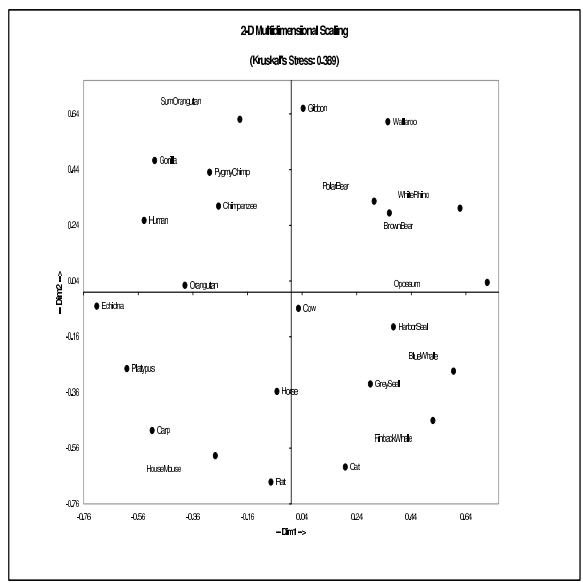,width=3in,height=3in,angle=270} \hfill\
\caption{Multidimensional clustering of same \NCD matrix
(Figure~\ref{fig.distmatr}) as used for Figure~\ref{tree-mammal}.
Kruskal's stress-1 = 0.389.
}\label{fig.mammal2d}
\end{figure}
In recent
years, as the complete genomes of various species become available,
it has become possible to do whole genome phylogeny (this overcomes the
problem that using different targeted parts
of the genome, or proteins, may give different trees \cite{RWKC03}).
Traditional phylogenetic methods on individual genes depended
on multiple alignment of the related proteins
and on the model of evolution of individual amino acids.
Neither of these is practically applicable to the genome level.
In absence of such models, a method which can compute the shared
information between two sequences is useful because biological
sequences encode information, and the occurrence of evolutionary
events (such as insertions, deletions, point mutations,
rearrangements, and inversions) separating two sequences sharing a
common ancestor will result in the loss of their shared
information. Our method (in the form of the CompLearn Toolkit) is
a fully automated
software tool based on such a distance to compare two
genomes.
In evolutionary biology the
timing and origin of the major extant placental clades
(groups of organisms that have evolved from a common ancestor)
continues to fuel debate and research. Here, we provide evidence
by whole mitochondrial genome phylogeny for competing hypotheses
in two main questions: the grouping of the Eutherian orders,
and the Therian hypothesis versus the Marsupionta hypothesis.

{\bf Eutherian Orders:}
We demonstrate (already in \cite{Li03}) that
a whole mitochondrial genome phylogeny of the Eutherians (placental
mammals)
can be reconstructed automatically from {\em
unaligned} complete mitochondrial genomes by use of an
early form of our compression method,
using standard software packages.
%We will use whole mitochondrial DNA genomes of 20 mammals and the
%problem of Eutherian orders to
%make a comprehensive examination of our measures.
%The problem we consider is this:
As more genomic material has become available,
the debate in biology has intensified
 concerning which two of the three main groups of placental
mammals are more closely related: Primates, Ferungulates, and Rodents.
In \cite{Cao1998}, the maximum likelihood method of phylogeny tree
reconstruction gave evidence for the
(Ferungulates, (Primates, Rodents)) grouping for half of
the proteins in the mitochondial genomes investigated, and
(Rodents, (Ferungulates, Primates)) for the other halves of the mt genomes.
In that experiment
they aligned 12 concatenated
mitochondrial proteins, taken from 20 species: rat ({\em Rattus
norvegicus}), house mouse ({\em Mus musculus}),
grey seal ({\em
Halichoerus grypus}), harbor seal ({\em Phoca vitulina}), cat ({\em
Felis catus}), white rhino ({\em Ceratotherium simum}), horse ({\em
Equus caballus}), finback whale ({\em Balaenoptera physalus}), blue
whale ({\em Balaenoptera musculus}), cow ({\em Bos taurus}), gibbon
({\em Hylobates lar}), gorilla ({\em Gorilla gorilla}), human ({\em
Homo sapiens}), chimpanzee ({\em Pan troglodytes}), pygmy chimpanzee
({\em Pan paniscus}), orangutan ({\em Pongo pygmaeus}), Sumatran
orangutan ({\em Pongo pygmaeus abelii}), using opossum ({\em Didelphis
virginiana}), wallaroo ({\em Macropus robustus}), and the platypus ({\em
Ornithorhynchus anatinus}) as outgroup.

{\bf Marsupionta and Theria:}
The extant monophyletic divisions of the class Mammalia are
the Prototheria (monotremes: mammals that procreate using eggs),
Metatheria (marsupials: mammals that procreate using pouches), and
Eutheria (placental mammals: mammals that procreate using placentas).
The sister relationships between
these groups is viewed as the most fundamental question in
mammalian evolution  \cite{KBSMJ01}. Phylogenetic comparison by either
anatomy or mitochondrial genome has resulted in two conflicting
hypotheses: the gene-isolation-supported {\em Marsupionta hypothesis}:
((Prototheria, Metatheria), Eutheria)
versus the morphology-supported {\em Theria hypothesis}:
(Prototheria, (Methateria, Eutheria)),
the third possiblity apparently not being held seriously by anyone.
There has been a lot of support for either hypothesis;
recent support for the Theria hypothesis was given
in \cite{KBSMJ01} by analyzing a large nuclear gene (M6P/IG2R),
viewed as important
across the species concerned, and even more recent support for
the Marsupionta hypothesis was given in \cite{JMWWA02} by
phylogenetic analysis of another sequence from the nuclear gene
(18S rRNA) and by the whole mitochondrial genome.

{\bf Experimental Evidence:} To test the Eutherian orders simultaneously
with the Marsupionta- versus Theria hypothesis,
we added four animals to the above twenty:
Australian
echidna ({\em Tachyglossus aculeatus}),
brown bear ({\em Ursus arctos}),
polar bear ({\em Ursus maritimus}),
using the common carp
({\em Cyprinus carpio}) as the outgroup.
Interestingly, while there are many species of Eutheria and Metatheria,
there are only three species of now living Prototheria known: platypus,
and two types of echidna (or spiny anteater). So our sample of the Prototheria
is large.
The whole mitochondrial genomes of the total of 24 species we used  were
downloaded from the
GenBank Database on the world-wide web. Each is around 17,000 bases.
The \NCD distance matrix was computed using the compressor PPMZ.
The resulting phylogeny, with an almost maximal $S(T)$ score
of 0.996 supports anew the currently
accepted grouping (Rodents, (Primates, Ferungulates)) of the Eutherian orders,
and additionally the Marsupionta hypothesis
((Prototheria, Metatheria), Eutheria),
see Figure \ref{tree-mammal}. Overall, our whole-mitochondrial
\NCD analysis supports the following hypothesis:

{\footnotesize{
\[
\rm
 \overbrace{(\underbrace{(primates, ferungulates)(rodents}_{\rm Eutheria},
 (Metatheria,
Prototheria)))}^{Mammalia},
\]
}}

which indicates that the rodents, and the branch leading to
the Metatheria and Prototheria, split off early from the branch that
led to the primates and ferungulates. Inspection of the distance matrix shows
that the primates are very close together, as are the rodents, the Metatheria,
and the Prototheria. These are tightly-knit groups with relatively close
\NCD's. The ferungulates are a much looser group with generally distant
\NCD's. The intergroup distances show that the Prototheria are
furthest away from the other groups, followed by the Metatheria and
the rodents.
Also the fine-structure of the tree is consistent with biological wisdom.
\begin{figure*}
\begin{center}
{\tiny
\begin{verbatim}
             BlueWhale            Cat             Echidna           Gorilla            Horse            Opossum          PolarBear         SumOrang
                   BrownBear         Chimpanzee         FinWhale          GreySeal        HouseMouse         Orangutan         PygmyChimp         Wallaroo
                           Carp               Cow              Gibbon          HarborSeal          Human            Platypus            Rat            WhiteRhino
     BlueWhale 0.005 0.906 0.943 0.897 0.925 0.883 0.936 0.616 0.928 0.931 0.901 0.898 0.896 0.926 0.920 0.936 0.928 0.929 0.907 0.930 0.927 0.929 0.925 0.902
     BrownBear 0.906 0.002 0.943 0.887 0.935 0.906 0.944 0.915 0.939 0.940 0.875 0.872 0.910 0.934 0.930 0.936 0.938 0.937 0.269 0.940 0.935 0.936 0.923 0.915
          Carp 0.943 0.943 0.006 0.946 0.954 0.947 0.955 0.952 0.951 0.957 0.949 0.950 0.952 0.956 0.946 0.956 0.953 0.954 0.945 0.960 0.950 0.953 0.942 0.960
           Cat 0.897 0.887 0.946 0.003 0.926 0.897 0.942 0.905 0.928 0.931 0.870 0.872 0.885 0.919 0.922 0.933 0.932 0.931 0.885 0.929 0.920 0.934 0.919 0.897
    Chimpanzee 0.925 0.935 0.954 0.926 0.006 0.926 0.948 0.926 0.849 0.731 0.925 0.922 0.921 0.943 0.667 0.943 0.841 0.946 0.931 0.441 0.933 0.835 0.934 0.930
           Cow 0.883 0.906 0.947 0.897 0.926 0.006 0.936 0.885 0.931 0.927 0.890 0.888 0.893 0.925 0.920 0.931 0.930 0.929 0.905 0.931 0.921 0.930 0.923 0.899
       Echidna 0.936 0.944 0.955 0.942 0.948 0.936 0.005 0.936 0.947 0.947 0.940 0.937 0.942 0.941 0.939 0.936 0.947 0.855 0.935 0.949 0.941 0.947 0.929 0.948
  FinbackWhale 0.616 0.915 0.952 0.905 0.926 0.885 0.936 0.005 0.930 0.931 0.911 0.908 0.901 0.933 0.922 0.936 0.933 0.934 0.910 0.932 0.928 0.932 0.927 0.902
        Gibbon 0.928 0.939 0.951 0.928 0.849 0.931 0.947 0.930 0.005 0.859 0.932 0.930 0.927 0.948 0.844 0.951 0.872 0.952 0.936 0.854 0.939 0.868 0.933 0.929
       Gorilla 0.931 0.940 0.957 0.931 0.731 0.927 0.947 0.931 0.859 0.006 0.927 0.929 0.924 0.944 0.737 0.944 0.835 0.943 0.928 0.732 0.938 0.836 0.934 0.929
      GreySeal 0.901 0.875 0.949 0.870 0.925 0.890 0.940 0.911 0.932 0.927 0.003 0.399 0.888 0.924 0.922 0.933 0.931 0.936 0.863 0.929 0.922 0.930 0.920 0.898
    HarborSeal 0.898 0.872 0.950 0.872 0.922 0.888 0.937 0.908 0.930 0.929 0.399 0.004 0.888 0.922 0.922 0.933 0.932 0.937 0.860 0.930 0.922 0.928 0.919 0.900
         Horse 0.896 0.910 0.952 0.885 0.921 0.893 0.942 0.901 0.927 0.924 0.888 0.888 0.003 0.928 0.913 0.937 0.923 0.936 0.903 0.923 0.912 0.924 0.924 0.848
    HouseMouse 0.926 0.934 0.956 0.919 0.943 0.925 0.941 0.933 0.948 0.944 0.924 0.922 0.928 0.006 0.932 0.923 0.944 0.930 0.924 0.942 0.860 0.945 0.921 0.928
         Human 0.920 0.930 0.946 0.922 0.667 0.920 0.939 0.922 0.844 0.737 0.922 0.922 0.913 0.932 0.005 0.949 0.834 0.949 0.931 0.681 0.938 0.826 0.934 0.929
       Opossum 0.936 0.936 0.956 0.933 0.943 0.931 0.936 0.936 0.951 0.944 0.933 0.933 0.937 0.923 0.949 0.006 0.960 0.938 0.939 0.954 0.941 0.960 0.891 0.952
     Orangutan 0.928 0.938 0.953 0.932 0.841 0.930 0.947 0.933 0.872 0.835 0.931 0.932 0.923 0.944 0.834 0.960 0.006 0.954 0.933 0.843 0.943 0.585 0.945 0.934
      Platypus 0.929 0.937 0.954 0.931 0.946 0.929 0.855 0.934 0.952 0.943 0.936 0.937 0.936 0.930 0.949 0.938 0.954 0.003 0.932 0.948 0.937 0.949 0.920 0.948
     PolarBear 0.907 0.269 0.945 0.885 0.931 0.905 0.935 0.910 0.936 0.928 0.863 0.860 0.903 0.924 0.931 0.939 0.933 0.932 0.002 0.942 0.940 0.936 0.927 0.917
    PygmyChimp 0.930 0.940 0.960 0.929 0.441 0.931 0.949 0.932 0.854 0.732 0.929 0.930 0.923 0.942 0.681 0.954 0.843 0.948 0.942 0.007 0.935 0.838 0.931 0.929
           Rat 0.927 0.935 0.950 0.920 0.933 0.921 0.941 0.928 0.939 0.938 0.922 0.922 0.912 0.860 0.938 0.941 0.943 0.937 0.940 0.935 0.006 0.939 0.922 0.922
  SumOrangutan 0.929 0.936 0.953 0.934 0.835 0.930 0.947 0.932 0.868 0.836 0.930 0.928 0.924 0.945 0.826 0.960 0.585 0.949 0.936 0.838 0.939 0.007 0.942 0.937
      Wallaroo 0.925 0.923 0.942 0.919 0.934 0.923 0.929 0.927 0.933 0.934 0.920 0.919 0.924 0.921 0.934 0.891 0.945 0.920 0.927 0.931 0.922 0.942 0.005 0.935
    WhiteRhino 0.902 0.915 0.960 0.897 0.930 0.899 0.948 0.902 0.929 0.929 0.898 0.900 0.848 0.928 0.929 0.952 0.934 0.948 0.917 0.929 0.922 0.937 0.935 0.002
\end{verbatim}
}
\caption{Distance matrix of pairwise \NCD. For display purpose,
we  have truncated the original
entries from 15 decimals to 3 decimals precision.}\label{fig.distmatr}
\end{center}
\end{figure*}

\section{Hierarchical versus Flat Clustering}
This is a good place to contrast the informativeness of
hierarchical clustering with multidimensional clustering
using the same \NCD matrix, exhibited in
Figure~\ref{fig.distmatr}. The entries give a good example of typical
\NCD values; we truncated the number of decimals
from 15 to 3 significant digits to save space.
Note that the majority of distances bunches in the range $[0.9,1]$. This is due
to the regularities the compressor can perceive.
The diagonal elements give the self-distance, which, for PPMZ, is not
actually 0, but is off from 0 only in the third decimal.
In Figure~\ref{fig.mammal2d} we clustered the 24 animals using the \NCD matrix
by multidimenional scaling as points in 2-dimensional Euclidean space.
In this method,
the \NCD matrix of 24 animals can be viewed as a set of
distances between points in $n$-dimensional Euclidean
space ($n \leq 24$), which
we want to project
into a 2-dimensional Euclidean space, trying to distort the distances
between the pairs
as little as possible. This is akin to the problem of projecting
the surface of the earth globe on a two-dimensional map with minimal
distance distortion. The main feature is the choice of the measure
of distortion to be minimized,  \cite{DHS}. Let the original set of distances
be $d_1, \ldots , d_k$ and the projected distances be
$d'_1, \ldots , d'_k$.
 In Figure~\ref{fig.mammal2d} we used
the distortion measure {\em Kruskall's stress-1},
\cite{Kr64}, which minimizes
$\sqrt{(\sum_{i \leq k} (d_i -d'_i)^2)/ \sum_{i \leq k} d_i^2}$.
Kruskall's stress-1 equal 0 means no distortion, and the worst value
is at most 1 (unless you have a really bad projection).
In the projection of the \NCD matrix according to our quartet method
one minimizes the more subtle distortion $S(T)$ measure, where 1 means
perfect representation of the relative relations between every 4-tuple,
and 0 means minimal representation.
Therefore, we should compare distortion Kruskall stress-1 with $1-S(T)$.
Figure~\ref{tree-mammal} has a very good
$1-S(T)=0.04$ and
Figure~\ref{fig.mammal2d} has a poor Kruskal stress $0.389$.
Assuming that the comparison is significant for small values (close to perfect
projection), we find that
the multidimensional scaling of this experiment's \NCD matrix
is formally inferior to that of the quartet
tree. This conclusion formally justifies
the impression conveyed by the figures on visual inspection.

\begin{small}

\end{small}


\begin{thebibliography}{99}



\bibitem{BGLVZ98}
C.H.~Bennett, P.~G\'acs, M. Li, P.M.B.~Vit\'anyi, and W.~Zurek.
Information Distance, {\em IEEE Transactions on Information Theory},
44:4(1998), 1407--1423.


\bibitem{BLM03}
C.H. Bennett, M. Li, B. Ma, Chain letters and evolutionary histories,
{\em Scientific American}, June 2003, 76--81.

\bibitem{BCGOP98}
A. Ben-Dor, B. Chor, D. Graur, R. Ophir, D. Pelleg,
Constructing phylogenies from quartets: Elucidation of eutherian
superordinal relationships, {\em J. Computational Biology},
5:3(1998), 377--390.

\bibitem{BJKLW99}
V. Berry, T. Jiang, P. Kearney, M. Li, T. Wareham,
Quartet cleaning: improved algorithms and simulations.
Algorithms--Proc. 7th European Symp. (ESA99),
LNCS vol. 1643, Springer Verlag, Berlin, (1999), 313-324.


\bibitem{Br00}
D.~Bryant, V.~Berry, P.~Kearney, M.~Li, T.~Jiang,
T.~Wareham and H.~Zhang. A practical algorithm for
recovering the best supported edges of an evolutionary tree.
{\em Proc. 11th  ACM-SIAM Symposium on Discrete Algorithms},
January 9--11, 2000,
San Francisco, California, USA,
287--296, 2000.

\bibitem{Bu71}
P. Buneman, The recovery of trees from measures of dissimilarity.
Pp. 387--395 in: F. Hodson, D. Kenadall, P. Tautu (Eds.),
{Proc. of the Anlo-Romanian conference}, The Royal Society of London and
the Academy of the Socialist Republic of Romania, The University Press,
Edinburgh, Scottland, UK.

\bibitem{Cao1998}
Y. Cao, A. Janke, P. J. Waddell, M. Westerman,
O. Takenaka, S. Murata, N. Okada, S. P\"a\"abo, M. Hasegawa,
Conflict among individual mitochondrial proteins in resolving the
phylogeny of Eutherian orders,
{\em J. Mol. Evol.}, 47(1998), 307-322.

\bibitem{SID}
X. Chen, B. Francia, M. Li, B. McKinnon, A. Seker,
Shared information and program plagiarism detection,
{\em IEEE Trans. Inform. Th.}, 50:7(2004), 1545--1551.

\bibitem{Ci03}
R. Cilibrasi,
The CompLearn Toolkit, 2003,
 http://complearn.sourceforge.net/   .

\bibitem{CVW03}
R. Cilibrasi, P.M.B. Vitanyi, R. de Wolf,
Algorithmic clustering of music based on string compression,
{\em Computer Music J.}, 28:4(2004), 49-67. Web archives version
http://xxx.lanl.gov/abs/cs.SD/0303025 of 24 March 2003.

\bibitem{CV04}
R. Cilibrasi, P.M.B. Vitanyi, Clustering by compression,
{\em IEEE Trans. Information Theory}, 51:4(2005), 1523- 1545.

\bibitem{CS77}
H. Colonius, H.H. Schulze, Trees constructed from empirical relations,
{\em Braunschweiger Berichte as dem Institut fuer Psychologie},
1(1977).

\bibitem{CS81}
H. Colonius, H.-H. Schulze, Tree structures for proximity data.
{\em British Journal of Mathematical and Statistical Psychology}, 34(1981),
167-180.

\bibitem{DHS}
R.O. Duda, P.E. Hart, D.G. Stork, {\em Pattern Classification},
2nd Edition, Wiley Interscience, 2001.

\bibitem{Fe81}
Felsenstein, J. Evolutionary trees from DNA sequences: a maximum likelihood approach.
{\em J. Molecular Evolution} 17(1981), 368--376.

\bibitem{JMWWA02}
A. Janke, O. Magnell, G. Wieczorek, M. Westerman, U. Arnason,
Phylogenetic analysis of 18S rRNA and the mitochondrial
genomes of wombat, Vombatus ursinus, and the spiny
anteater, Tachyglossus acelaetus: increased support for
the Marsupionta hypothesis,
{\em J. Mol. Evol.}, 1:54(2002), 71--80.

\bibitem{Ji01}
T.~Jiang, P.~Kearney, and M.~Li.
A Polynomial Time Approximation Scheme for Inferring Evolutionary Trees from
Quartet Topologies and its Application.
{\em SIAM J. Computing}, 30:6(2000), 1942--1961.

\bibitem{Ke04}
E. Keogh, S. Lonardi, and C.A. Rtanamahatana, Toward parameter-free
data mining, In: {\em Proc. 10th ACM SIGKDD Intn'l Conf. Knowledge
Discovery and Data Mining}, Seattle, Washington, USA, August 22---25, 2004,
206--215.

\bibitem{KBSMJ01}
J.K. Killian, T.R. Buckley, N. Steward, B.L. Munday, R.L. Jirtle,
Marsupials and Eutherians reunited: genetic evidence for the Theria
hypothesis of mammalian evolution, {\em Mammalian Genome}, 12(2001),
513--517.

\bibitem{KGV83}
S. Kirkpatrick, C.D. Gelatt and M.P. Vecchi, {\em Science} 220 (1983) 671-680.


\bibitem{KSAG03}
A. Kraskov, H. St\"ogbauer, R.G. Adrsejak, P. Grassberger,
Hierarchical clustering based on mutual information, 2003,
http://arxiv.org/abs/q-bio/0311039

\bibitem{Kr64}
J.B. Kruskal,
Nonmetric multidimensional scaling: a numerical method, {\em Psychometrika},
29(1964), 115--129.


\bibitem{SA03}
T.G. Ksiazek, et.al.,
A Novel Coronavirus Associated with Severe Acute Respiratory Syndrome,
{\em New England J. Medicine}, Published at www.nejm.org April 10,
2003 (10.1056/NEJMoa030781).

%\bibitem{KH98}
%S. Kumar, B. Hodges, A molecular timescale for vertebrate evolution,
%{\em Nature}, 392(1998), 30 April, 917--919.


\bibitem{Ko89}
J.R. Koza,
Hierarchical genetic algorithms operating on
                 populations of computer programs,
{\em Proc. 11th Intn'l Joint Conf. Artificial Intell. (IJCAI'89)},
Morgan-Kaufmann, 1989, 768--774.


\bibitem{La51}
P.S. Laplace, {\it A philosophical essay on probabilities}, 1819.
English translation, Dover, 1951.


\bibitem{LBCKKZ01}
M.~Li, J.H.~Badger, X.~Chen, S.~Kwong, P.~Kearney, and H.~Zhang.
An information-based sequence distance and its application
to whole mitochondrial genome phylogeny,
{\em Bioinformatics}, 17:2(2001), 149--154.

\bibitem{Li01}
M.~Li and P.M.B.~Vit\'anyi.
Algorithmic Complexity,
pp.~376--382 in: {\em International Encyclopedia
of the Social \& Behavioral Sciences},
N.J.~Smelser and P.B.~Baltes, Eds., Pergamon, Oxford, 2001/2002.

\bibitem{Li03}
M. Li, X. Chen, X.~Li, B.~Ma, P.M.B.~Vit\'anyi.
The similarity metric, {\em IEEE Trans. Inform. Th.}, 50:12(2004),
3250- 3264.

\bibitem{LiVi97}
M.~Li and P.M.B.~Vit\'anyi.
{\em An Introduction to Kolmogorov Complexity
and its Applications}, Springer-Verlag, New York, 2nd Edition, 1997.

\bibitem{LTM05}
T. Liu, J. Tang, B.M.E. Moret,
Quartet methods for phylogeny reconstruction from gene orders.
Dept. CS and Engin., Univ. South-Carolina, Columbia,
SC, USA. Manuscript.

\bibitem{MRRTT53}
N. Metropolis, A.W. Rosenbluth, M.N. Rosenbluth.
A.H. Teller and E. Teller, {\em J. Chem. Phys.} 21 (1953) 1087-1092.

\bibitem{Ni06}
H.~Niman,
Recombinomics website, 2006,
 http://www.recombinomics.com/   .

\bibitem{PBE04}
R. Piaggio-Talice, J. Gordon Burleigh, O. Eulenstein,
Quartet supertrees. Chapter 4, pp. 173--191 in:
O.R.P. Beninda-Edmonds (Ed.), {\em Phylogenetic Supertrees:
Combining Information to Reveal the Tree of Life}. Computational Biology,
volume 3 (Dress. A., series ed.), Kluwer Academic Publishers, 2004.

\bibitem{RWKC03}
A. Rokas, B.L. Williams, N. King, S.B. Carroll,
Genome-scale approaches to resolving incongruence in molecular
phylogenies, {\em Nature}, 425(2003), 798--804 (25 October 2003).

\bibitem{RMWW04}
U. Roshan, B.M.E. Moret, T. Warnow, T.L. Williams,
Performance of supertree methods on various  datasets decompositions,
pp 301--328 in:
O.R.P. Beninda-Edmonds (Ed.), {\em Phylogenetic Supertrees:
Combining Information to Reveal the Tree of Life}. Computational Biology,
volume 3 (Dress. A., series ed.), Kluwer Academic Publishers, 2004.


\bibitem{SN87}
N. Saitou, M. Nei, The neighbor-joining method: a new method for
reconstructing phylogenetic trees, {\em Mol. Biol. Evol.}, 4(1987), 406--425.


\bibitem{St92}
M. Steel, The complexity of reconstructiong trees form qualitative
characters and subtrees, {\em Journal of Classification},
9(1992), 91--116.

\bibitem{SvH96}
K. Strimmer, A. von Haeseler, Quartet puzzling: A quartet maximum-likelihood
method for reconstructing tree topologies, {\em Mol. Biol. Evol.},
13:7(1996), 964--969.

\bibitem{Vi00}
P.M.B. Vitanyi, A discipline of evolutionary programming,
{\em Theoret. Comp. Sci.}, 241:1-2 (2000), 3--23.

\bibitem{We05}
S. Wehner, Analyzing network traffic and worms using compression,
http://arxiv.org/pdf/cs.CR/0504045

\bibitem{WDGG05}
J. Weyer-Menkoff, C. Devauchelle, A. Grossmann, S. Gr\"unewald,
Integer linear programming as a tool for constructing trees from
quartet data. Preprint from the web submitted to Elsevier Science.


\end{thebibliography}
\end{document}